\documentclass[reprint,amsmath,amssymb,aps,nofootinbib]{revtex4-1}

\usepackage{graphicx}
\usepackage{dcolumn}
\usepackage{bm}
\usepackage[normalem]{ulem}

\usepackage{color}
\usepackage{xcolor}

\usepackage{mathrsfs}
\usepackage{amsfonts}
\usepackage{varioref}
\usepackage{csquotes}
\usepackage{hyphenat}
\usepackage{float}
\usepackage{multirow}

\RequirePackage[colorlinks,citecolor=blue,urlcolor=blue,linkcolor=blue]{hyperref}

\def\eq#1{{Eq.~(\ref{#1})}}

\def\sect#1{{Sec.~\ref{#1}}}

\definecolor{darkpastelgreen}{rgb}{0.01, 0.75, 0.24}

\begin{document}

\title{Resonance crossing of a charged body in a magnetized Kerr background:\\ an analogue of extreme mass ratio inspiral}

\author{Sajal Mukherjee$^1$}
\email{mukherjee@asu.cas.cz}
\author{Ond\v{r}ej Kop\'{a}\v{c}ek$^{1,2}$}
\email{kopacek@asu.cas.cz}
\author{Georgios Lukes-Gerakopoulos$^1$}
\email{gglukes@gmail.com}
\affiliation{${}^1$ Astronomical Institute of the Czech Academy of Sciences, Bo\v{c}n\'{i} II 1401/1a, CZ-141 00 Prague, Czech Republic}  
\affiliation{${}^2$Faculty of Science, Humanities and Education, Technical University of Liberec,  Studentsk\'{a} 1402/2, CZ-461\,17~Liberec, Czech Republic}

\begin{abstract}
\noindent
We investigate resonance crossings of a charged body moving around a Kerr black hole immersed in an external homogeneous magnetic field. This system can serve as an electromagnetic analogue of a weakly non-integrable extreme mass ratio inspiral (EMRI). In particular, the presence of the magnetic field renders the conservative part of the system non-integrable in the  Liouville sense, while the electromagnetic self-force causes the charged body to inspiral. By studying the system without the self-force, we show the existence of an approximate Carter-like constant and discuss how resonances grow as a function of the perturbation parameter. Then, we apply the electromagnetic self-force to investigate crossings of these resonances during an inspiral. Averaging the energy and angular momentum losses during crossings allows us to employ an adiabatic approximation for them. We demonstrate that such adiabatic approximation provides results qualitatively equivalent to the instantaneous self-force evolution, which indicates that the adiabatic approximation may describe the resonance crossing with sufficiently accuracy in EMRIs. 

\end{abstract}

\maketitle

\section{Introduction}

Dynamical systems under non-integrable perturbation exhibit various extraordinary features compared to the unperturbed ones \citep{lichtenberg92}. First of all, the perturbation may trigger chaotic motions in some regions of the phase space. Depending on the strength of the perturbation, deterministic chaos may completely dominate the dynamics. However, even in the slightly perturbed systems with negligible amount of chaos, we observe dynamically relevant non-integrable effects near resonances, i.e. in parts of the phase space, where the characteristic frequencies of the system are commensurate. In fact, according to the {\em Kolmogorov-Arnold-Moser} (KAM) theorem \cite{Arnold63}, parts of the phase space of a weakly perturbed system that are far enough from resonances remain basically unaffected  by the imposed perturbation. On the other hand, the dynamics in the vicinity of resonances considerably differs which may affect measurable properties of the system. Motivated by the latter fact, there were studies trying to provide some insight into the dynamics of an Extreme Mass Ratio Inspiral (EMRI) crossing the resonance \cite{Apostolatos09,LGAC10,Destounis20,Destounis21,LG21,Bronicki22}. 

EMRIs represent a key observational target for the future mission of Laser Interferometer Space Antenna (LISA) \cite{EMRIsLISA}. They are composed of a supermassive primary black hole and a much lighter secondary compact object (black hole or neutron star) inspiralling into the primary.  The mass ratio $\eta=m/M$ between the mass $m$ of the secondary to the mass $M$ of the primary is smaller than $10^{-4}$. The smallness of $\eta$ allows us to  use it as a perturbation parameter when expanding the background spacetime of this binary system in orders of $\eta$ to calculate the gravitational self-force (GSF) \cite{Barack19,Pound21}. Actually, the dissipative part of this self-force causes the secondary to inspiral towards the primary as it radiates away energy and angular momentum in the form of gravitational waves.

The full first order self-force for a non-spinning secondary moving on a generic orbit has been obtained relatively recently  \cite{vandeMeent18,Pound21}, but using the full self-force is computationally expensive. The main cause is that the inspiralling body revolves around the primary body for a number of cycles inversely proportional to the mass ratio of the system ($\eta^{-1}$). Since the mass ratio in an EMRI is small, the number of cycles becomes very large making the numeric computations highly demanding. In order to decrease the computational demands, the gravitational self-force may be approximated in some way. In particular, we may employ an adiabatic approximation \cite{Drasco06}, which takes into account just the averaged dissipative part of the first order self-force, while more advanced approximations are also available \cite{Isoyama21,Katz21}.

Actually, Ref.~\cite{Isoyama21} attempts to tackle the issue that during the many EMRI cycles, it is almost certain that the inspiralling body will cross resonances, i.e. some of the orbital characteristic frequencies of the secondary will become commensurate \cite{Apostolatos09,Flanagan12,Brink15,Brink:2015roa,Mukherjee:2019jhd}. Not all of these crossings are equally important. Only those with a small denominator, like $1:2$, $2:3$, are expected to have a significant impact on the inspiral \cite{Berry16}. It has not yet been clarified up to which value of the denominator we should expect this impact, but we speculate that it should be a value of the order of $10$. Most of the models we use to approximate an EMRI are problematic at the resonances \cite{Berry16}, however, as was already mentioned there is an ongoing effort to overcome this obstacle \cite{Isoyama21,LG21} and this work is part of this effort.

Mutual effects of the electromagnetic and the gravitational self-force in the dynamics of a charged compact body has recently been studied in the context of EMRIs \citep{sarkar21}. That study has shown that besides the interaction terms previously derived in \cite{Zimmerman:2014uja}, additional perturbative terms linear in the metric perturbation are generated and these terms may become relevant in some astrophysical situations. However, here we adopt a different approach and we do not consider combined effects of the both interactions. In our scenario, the electromagnetic self-force exerted on the charged body in a magnetized spacetime serves as an analogue model of a gravitational self-force \citep{poisson04} which we, however, do not consider explicitly. Although, the evolution of an EMRI system and the emission of the gravitational radiation are actually governed by the gravitational self-force, here we deal with the aspects of the dynamics correlated with the resonances and their crossing due to dissipation within the electromagnetic analogue framework, which allows us to avoid the intricate formalism of the gravitational self-force \cite{Pound05,Pound08}.
 
The primary motivation of this work is to study the system with the electromagnetic self-force as an analogue of an EMRI, however, the fact that we study dynamics of an electrically charged body in the vicinity of a magnetized rotating black hole, makes the results of our analysis relevant also to other astrophysical applications. Hence, let us discuss a little bit the framework of this model. We consider an asymptotically uniform magnetic field aligned with the rotation axis of the black hole described by a vacuum model derived by Wald in \citep{wald74}. Albeit asymptotically uniform, the field becomes largely deformed due to frame-dragging and other relativistic effects if we get closer to the horizon of the black hole \citep{bicak85, karas13}. Although the employed weak-field approximation does not take into account the effect of the field onto the curvature of the spacetime, the geometric effects of the spinning black hole onto the topology of the electromagnetic field are described completely by the given model.

On the other hand, the effect of the charged matter onto the electromagnetic field is neglected in this framework. This might appear as an issue especially within the inner parts of the accretion disk where the presence of charges and currents significantly contributes to the field and, in particular, small-scale  magnetic fields may be induced here due to turbulence driven by the magnetorotational instability \citep{balbus98}. Turbulence allows the transport of the angular momentum and, thus, contributes to the viscosity of the disk which is crucial for the accretion process. Nevertheless, for the study of the resonance crossing, the small-scale structure of the magnetic field and full description of the physics of accretion is not relevant. In fact, the organized large-scale field \citep{wald74} provides an appropriate framework allowing the dissipation of the energy and angular momentum due to radiation losses of the charged test-body \citep{kolos21}.

The model of an axisymmetric vacuum magnetosphere of the rotating black hole described by the Wald's solution \citep{wald74} of the Maxwell's equations on the Kerr background has been employed in various context. For instance, it has been shown that this model (unlike the pure Kerr or Kerr-Newman background) allows stable off-equatorial circular orbits \citep{kovar10}, which are astrophysically relevant as a basic model  to study the dynamics of diluted plasma in an accretion disk coronae. Moreover, the magnetic field acts as a non-integrable perturbation triggering (deterministic) chaos in some regions of the phase space \citep{kopacek10} and chaotic dynamics may contribute to launching the outflow of escaping jet-like trajectories \citep{alzahrani14,kopacek18b,kopacek20}. Hence, although, the model of vacuum magnetosphere does not attempt to provide a complete description of the field topology of an accreting black hole, it represents an useful approximation which allows the study of various astrophysically relevant processes. 

The rest of the article is organized as follows. Sec.~\ref{sec:ChargedPartIntro} introduces the system of a charged body moving in a Kerr background with a test external magnetic field without self-force. Sec.~\ref{sec:ResGrow} discusses how a resonances grow in this system. Sec.~\ref{sec:EMsf} studies the crossings of resonances when the dissipation due to instantaneous electromagnetic self-force is imposed, while Sec.~\ref{sec:adiabatic} compares the instantaneous self-force results with the adiabatic ones. Finally, Sec.~\ref{sec:Param} examines our parameter choices and Sec.~\ref{sec:Conc} discusses our main findings.

\section{Motion of a charged body in an external electromagnetic field}
\label{sec:ChargedPartIntro}

In this section, we discuss the motion of a charged body around a magnetized Kerr black hole without the electromagnetic self-force. We start with the equations of motion and the conserved quantities, and then we discuss the integrability of the system and the existence of a Carter-like constant.

\subsection{Equations of motion}
The equations of motion for a charged body in a curved background read 
\begin{equation}
    \dfrac{D\mathcal{U}^{\mu}}{d\tau}=\tilde{q}F^{\mu}_{~\nu}~\mathcal{U}^{\nu},\label{eq:LORENTZ}
\end{equation}
where, \enquote*{D} denotes a covariant derivative, $\tau$ is the proper time, $\tilde{q}=q/m$ denotes the specific charge of a body with rest mass $m$, $\mathcal{U}$ is the 4-velocity, and $F_{\mu\nu}$ is the electromagnetic field tensor. 

The equations of motion \eqref{eq:LORENTZ} actually represent a set of second order differential equations in Boyer-Lindquist coordinates $x^{\mu}=(t,r,\theta,\phi)$. However, one can employ the Hamiltonian formalism to obtain equivalent set of first order equations for the canonical coordinates $(x^{\mu},\pi_{\nu})$, where $\pi_{\nu} =(\pi_t,\pi_r,\pi_{\theta},\pi_{\phi})$ are components of the canonical four-momentum. The Hamiltonian of a body  with electric charge $q$ and rest mass $m$ in  a vector potential field $A_{\mu}$ can be defined as \citep{mtw}:
\begin{equation}
\label{eq:hamiltonian}
H=\textstyle{\frac{1}{2}}g^{\mu\nu}(\pi_{\mu}-qA_{\mu})(\pi_{\nu}-qA_{\nu}),
\end{equation}
where $g_{\mu\nu}$ is the metric of the background spacetime. The vector potential $A_{\mu}$ is related to the electromagnetic field tensor as $F_{\mu \nu}=\partial_{\mu}A_{\nu}-\partial_{\nu}A_{\mu}$.

The equations of motion are:
\begin{equation}
\label{hameq}
\frac{{\rm d}x^{\mu}}{{\rm d}\lambda}\equiv P^{\mu}=
\frac{\partial H}{\partial \pi_{\mu}},
\quad 
\frac{d\pi_{\mu}}{d\lambda}=-\frac{\partial H}{\partial x^{\mu}},
\end{equation}
where $\lambda\equiv\tau/m$ is a dimensionless affine parameter. By employing the first equation we obtain the kinematical four-momentum $P^{\mu}=\pi^{\mu}-qA^{\mu}$, and the conserved value of the Hamiltonian is therefore $H=-m^2/2$.

\subsection{Kerr spacetime}

Within this study we consider a fixed spacetime background of a rotating black hole. The line element of the Kerr metric in Boyer-Lindquist coordinates reads
\begin{align}
    d s^2 = g_{tt}~d t^2+2~g_{t\phi}~d t~d \phi + g_{\phi\phi}~d \phi^2 
       + g_{rr}~d r^2+g_{\theta\theta}~d \theta^2 \, , \label{eq:LinEl}
\end{align}
where
 \begin{eqnarray}\label{eq:KerrMetric}\
   g_{tt} &=&-\left(1-\frac{2 M r}{\Sigma}\right) \, ,\quad
   g_{t\phi} = -\frac{2 a M r \sin^2{\theta}}{\Sigma} \, ,\nonumber\\
   g_{\phi\phi} &=& \frac{(\varpi^4-a^2\Delta \sin^2\theta) \sin^2{\theta}}{\Sigma} \, ,\quad 
   g_{rr} = \frac{\Sigma}{\Delta} \, ,\\
   g_{\theta\theta} &=& \Sigma, \nonumber
 \end{eqnarray} 
with
 \begin{align}
  \Sigma = r^2+ a^2 \cos^2{\theta} \, ,\:
  \Delta = \varpi^2-2 M r \, ,\:
  \varpi^2 = r^2+a^2 \, , \label{eq:Kerrfunc} 
 \end{align}
where $M$ is the mass of the black hole and $a$ is the Kerr spin parameter.

\subsection{Asymptotically uniform magnetic field}
We employ a simple model of a vacuum magnetosphere consisting of an asymptotically uniform magnetic field aligned with the spin of the black hole. The relevant test-field solution of Maxwell's equations on the Kerr background may be derived exploiting the fact that in this case the Killing vectors themselves, as well as their linear combinations, solve the Maxwell's equation \citep{wald74}. 
In particular, the vector potential $A_{\mu}=(A_t,0,0,A_{\phi})$ of the solution corresponding to the magnetic field of the asymptotic strength $B_0$ may be expressed in terms of covariant components of the Kerr metric tensor $g_{\mu\nu}$ as follows:
\begin{equation}
 A_t=\frac{B_0}{2}\left(g_{t\phi}+2ag_{tt}\right),\,\,A_{\phi}=\frac{B_0}{2}\left(g_{\phi\phi}+2ag_{t\phi}\right).    
\label{wald}
\end{equation}

Since the vector potential of a stationary axisymmetric field does not depend on the coordinates $t$ and $\phi$, we obtain the following set of non-zero components of the electromagnetic field tensor:
\begin{eqnarray}
F_{\theta t}=&-&F_{t\theta}=\frac{2 B_0 M a r \tilde{x}\tilde{y} \left(a^2 - r^2\right)}{\Sigma^2},\\
F_{r t}=&-&F_{t r}=\frac{B_0 M a \left( \tilde{x}^2 - 2\right) \left(r^2-a^2 \tilde{y}^2\right)}{\Sigma^2},\\
\nonumber F_{r\phi}=&-&F_{\phi r}=\frac{B_0   \tilde{x}^2}{\Sigma^2}  (a^4  r   \tilde{x}^4 - 2  a^4  r   \tilde{x}^2 + a^4  r-   M  a^4   \tilde{x}^4\\ 
\nonumber&+& 3  M  a^4   \tilde{x}^2 - 2  M  a^4 - 2  a^2  r^3   \tilde{x}^2 + 2  a^2  r^3\\
&-&   M  a^2  r^2   \tilde{x}^2 + 2  M  a^2  r^2 + r^5),\\
\nonumber F_{\theta\phi}=&-&F_{\phi\theta}=\frac{B_0   \tilde{y}   \tilde{x}}{\Sigma^2} (a^6   \tilde{y}^4 + a^4  r^2   \tilde{y}^4 + 2  a^4  r^2   \tilde{y}^2\\
\nonumber &-& 2  M  a^4  r   \tilde{y}^4 - 2  M  a^4  r + 2  a^2  r^4   \tilde{y}^2 + a^2  r^4\\
&-& 4  M  a^2  r^3   \tilde{y}^2 + r^6),
\end{eqnarray}
where we set $\tilde{x}\equiv\sin\theta$ and $\tilde{y}\equiv\cos\theta$.

We consider the model of a {\it weakly magnetized} black hole, in which the contribution of the electromagnetic field to the stress-energy tensor $T^{\mu\nu}$ is neglected and the field, thus, does not affect the spacetime geometry nor the motion of electrically neutral bodies. Using such test-field approximation is justified in astrophysical conditions, while the relevant field intensities encountered in cosmic environments are too low to affect the geometry significantly even in extreme cases of neutron stars and magnetars \citep{beskin15}.

Even if the field given by \eq{wald} is asymptotically uniform, in the vicinity of a rotating black hole the field structure becomes distorted by the frame-dragging and other effects of strong gravity \citep[e.g., ][]{karas13}. In particular, the horizon of the Kerr black hole tends to expel the magnetic field as the spin increases. With the extreme rotation ($a=M$), the expulsion becomes complete and the invariant magnetic flux through each hemisphere of the horizon drops to zero in what is known as black hole Meissner effect \citep{bicak85,bicak00,Gurlebeck17}. While this effect only operates in an axisymmetric field \citep{Gurlebeck18}, other remarkable effects like the emergence of magnetic null points may arise in non-axisymmetric vacuum magnetospheres of black holes or neutron stars \citep{kopacek18,karas12}. Nevertheless, in the present paper we restrict the discussion to the perfectly axisymmetric model of Kerr black hole immersed into a weak asymptotically uniform magnetic field aligned with the spin axis described by \eq{wald}. 

\subsection{Conserved quantities} \label{sec:ConQuant}

In this section, we consider the system without the electromagnetic self-force. Therefore, the value of the Hamiltonian given by \eq{eq:hamiltonian} remains conserved as $H=-m^2/2$. Due to the stationarity and the axisymmetry of the system, the relevant components of the canonical four-momentum $\pi_t$ and $\pi_\phi$ are also conserved and define the integrals $E$ (energy) and $L_{\rm z}$ (axial component of the angular momentum) as follows:
\begin{eqnarray}\label{eq:PT&PPHI}
-E\equiv\pi_{t}=P_t+q A_t, \quad L_{\rm z}\equiv\pi_{\phi}=P_{\phi}+q A_{\phi}.
\end{eqnarray}
In addition to the timelike and spacelike Killing vectors, the Kerr spacetime is also endowed with a Killing tensor. Due to this, once we switch off the magnetic field (or consider electrically neutral bodies which are not affected by the weak field), we find the fourth conserved quantity, namely the Carter constant \citep{carter68}. The existence of four independent and in involution integrals of motion in a system of four degrees of freedom assures full Liouville integrability of the system \citep{lichtenberg92}. 

Below we discuss the existence of a Carter-like constant in the presence of the external magnetic field. In order to investigate that, we employ the Carter's theorem \cite{carter1973, Mukherjee:2015oaa}. This theorem assumes an axisymmetric spacetime with metric coefficients which may be expressed as:
\begin{eqnarray}
g^{tt}&=&(\mathcal{T}_{tt}+\Theta_{tt})/(\Sigma_r+\Sigma_{\theta}), g^{rr}=\mathcal{T}_{rr}/(\Sigma_r+\Sigma_{\theta}),  \nonumber \\
g^{\theta \theta}&=&\Theta_{\theta \theta}/(\Sigma_r+\Sigma_{\theta}), g^{t\phi}=(\mathcal{T}_{t\phi}+\Theta_{t\phi})/(\Sigma_r+\Sigma_{\theta}), \nonumber \\
g^{\phi \phi}&=&(\mathcal{T}_{\phi \phi}+\Theta_{\phi \phi})/(\Sigma_r+\Sigma_{\theta}),
\label{eq:conmetric}
\end{eqnarray}
where $\mathcal{T}_{tt}$, $\mathcal{T}_{t\phi}$, $\mathcal{T}_{\phi \phi}$, $\mathcal{T}_{rr}$ and $\Sigma_r$ are functions that only depend on the radial coordinate $r$, while $\Theta_{tt}$, $\Theta_{\theta \theta}$, $\Theta_{t\phi}$, $\Theta_{\phi \phi}$ and $\Sigma_{\theta}$ are functions of $\theta$ only. In such spacetime, if we can express a Hamiltonian $H_{\rm c}$ in the form:
\begin{equation} \label{eq:HamCarForm}
    H_{\rm c}=\dfrac{1}{2\,m}\dfrac{H_r+H_{\theta}}{\mathcal{W}_r+\mathcal{W}_{\theta}},
\end{equation}
where $H_r$ and $\mathcal{W}_r$ are functions of $r$ only, and $H_{\theta}$ and $\mathcal{W}_{\theta}$ are functions of $\theta$ only, then the following quantity commutes with the Hamiltonian:
\begin{equation}
    \mathcal{K}=2 \mathcal{W}_r H_{\rm c}-H_{r}=H_{\theta}-2 \mathcal{W}_{\theta} H_{\rm c}=\dfrac{\mathcal{W}_r H_{\theta}-\mathcal{W}_{\theta}H_r}{\mathcal{W}_{r}+\mathcal{W}_{\theta}}.
    \label{eq:Carter_expression}
\end{equation}
Given that our work concerns the Kerr background, by comparing the expressions in \eq{eq:conmetric} with the Kerr metric coefficients, we arrive at:
\begin{eqnarray}
&&\mathcal{T}_{tt}=-(r^2+a^2)^2/\Delta, \mathcal{T}_{t\phi}=-a(r^2+a^2)/\Delta, \mathcal{T}_{\phi \phi}=-a^2/\Delta \nonumber \\
&& \mathcal{T}_{rr}=\Delta, \mathcal{T}_{\theta \theta}=1, \Theta_{tt}=a^2 \sin^2\theta, \Theta_{\theta \theta}=1, \Theta_{t\phi}=a,\nonumber \\
&& \Theta_{\phi \phi}=1/\sin^2\theta, \Sigma_r=r^2, \Sigma_{\theta}=r^2 \cos^2\theta \, .
\end{eqnarray}

To check if we can write the Hamiltonian~\eqref{eq:hamiltonian} in a manner similar to \eq{eq:HamCarForm}, we first expand it:
\begin{align}
2H=g^{\mu \nu}P_{\mu}P_{\nu}&=g^{tt}(P_{t})^2+2g^{t\phi}P_{t}P_{\phi}+g^{\phi \phi}(P_{\phi})^2\nonumber \\
&+g^{rr}(P_r)^2+g^{\theta \theta}(P_{\rm \theta})^2\, .
\end{align}
In the above, the 4-momentum is related with the 4-velocity as $P_{\mu}=m \mathcal{U}_{\mu}$. Note that $P^{r}$ and $P^{\theta}$ are arbitrary, while $P_{t}$ and $P_{\phi}$ can be obtained from the relations given in Eq.~\eqref{eq:PT&PPHI}:
\begin{align}
P_{t} &=-E-(\epsilon/2)(g_{t\phi}+2ag_{tt})\, , \nonumber \\
P_{\phi} &=L_{\rm z}-(\epsilon/2)(g_{\phi \phi}+2ag_{t\phi}),    
\end{align}
where $\epsilon=qB_0$ is a key parameter which scales the electromagnetic interaction and introduces the non-integrable perturbation to the geodesic motion. Note that as we employ the weak-field solution given by \eq{wald}, the charge $q$ and the asymptotic magnetic induction $B_0$ always couple as $qB_0$ leaving $\epsilon$ as the only independent parameter.

Having done the above two steps allow us to rewrite the Hamiltonian as follows:
\begin{align}
2H &=g^{tt}E^2+2g^{t\phi}EL_{\rm z}+g^{\phi \phi}L^2_{\rm z}+g^{rr}(P_r)^2+g^{\theta \theta}(P_{\rm \theta})^2+\nonumber\\
 &+\epsilon E \left[g^{tt}(g_{t\phi}+2ag_{tt})+g^{t\phi}(g_{\phi \phi}+2ag_{t\phi})\right] \nonumber \\
 &-\epsilon L_{\rm z}\Big[g^{\phi \phi}(g_{\phi \phi}+2ag_{t\phi})+g^{t\phi}(g_{t\phi}+2ag_{tt})\Big]+\epsilon^2 \mathcal{F},
\end{align}
where we define $\mathcal{F}$ as:
\begin{eqnarray}
    \mathcal{F}=\dfrac{1}{4}\big[g^{tt}(g_{t\phi}+2 a g_{tt})^2+g^{\phi \phi}(g_{\phi \phi}+2 a g_{t\phi})^2+ \nonumber\\
    2 g^{t\phi}(g_{t\phi}+ 2 a g_{tt})(g_{\phi \phi}+2 a g_{t\phi})\big].
\end{eqnarray}
 The Hamiltonian can be written in a more compact form as follows:
\begin{equation}
    H=H_0-\dfrac{\epsilon}{2}\Big(L_{\rm z}-2aE\Big)+\epsilon^2 \mathcal{F}=H_{\rm int}+\epsilon^2 \mathcal{F},
    \label{eq:Hamiltonian_mag}
\end{equation}
where $H_0$ and $H_{\rm int}$ are defined as:
\begin{align} \label{eq:H0or}
H_{\rm int} &= H_0-\dfrac{\epsilon}{2}\Big(L_{\rm z}-2aE\Big), \\
H_0 &=\dfrac{1}{2}\Bigl(g^{tt}E^2-2g^{t\phi}EL_{\rm z}+g^{\phi \phi}L^2_{\rm z}+g^{rr}P^2_r+g^{\theta \theta}P^2_{\theta}\Bigr)\nonumber\\
&=\dfrac{1}{2\Sigma}\Bigl[\Bigl(\mathcal{T}_{tt}E^2-2\mathcal{T}_{t\phi}EL_{\rm z}+\mathcal{T}_{\phi \phi}L^2_{\rm z}+\Delta P^2_{\rm r}\Bigr)\nonumber \\ 
&+\Bigl(\Theta_{tt}E^2-2\Theta_{t\phi}EL_{\rm z}+\Theta_{\phi \phi}L^2_{\rm z}+P^2_{\theta}\Bigr)\Bigr]\, .
\end{align}
In the above, $H_0$ corresponds to the Hamiltonian for a Kerr geodesic. Interestingly, the linear order perturbation in $\epsilon$ is only introducing an additional constant term to this Hamiltonian, as can be seen from \eq{eq:Hamiltonian_mag}. Therefore, if the Hamiltonian $H_0$ is separable in $r$ and $\theta$, so should be $H_{\rm int}$. On the other hand, $\mathcal{F}$ is a coupling term in $r$ and $\theta$, which renders the total Hamiltonian $H$ not separable in these coordinates. Therefore, to be able to employ Carter's theorem, we drop the term $\mathcal{F}$, which linearizes $H$ in $\epsilon$, and work with $H_{\rm int}$.  By using the metric components given in \eq{eq:conmetric}, we can break \eq{eq:H0or} into two pieces:
\begin{eqnarray}
&&H^0_{r}=\Delta P^2_{\rm r}+\mathcal{T}_{tt}E^2-2\mathcal{T}_{t\phi}EL_{\rm z}+\mathcal{T}_{\phi \phi}L^2_{\rm z}, \nonumber \\ &&H^0_{\theta}=P^2_{\theta}+\Theta_{tt}E^2-2\Theta_{t\phi}EL_{\rm z}+\Theta_{\phi \phi}L^2_{\rm z},\,
\end{eqnarray}
and the Hamiltonian given in \eq{eq:Hamiltonian_mag} now reads
\begin{eqnarray}
  H_{\rm int}&=&\dfrac{1}{2\Sigma} \Big(H^0_{r}+H^0_{\theta}-\Sigma H_1\Big) \nonumber  \\
  &=&\dfrac{1}{2\Sigma}\Big(H^0_{r}-r^2 H_1+H^0_{\theta}-a^2 \cos^2\theta H_1\Big)\, ,  
\end{eqnarray}
where, $H_1=\epsilon(L_{\rm z}-2aE)$. By rewriting the above with $H_{r}=H^0_{r}-r^2 H_1$, and $H_{\theta}=H^0_{\theta}-a^2 \cos^2\theta H_1$, we obtain finally
\begin{eqnarray}
H_{\rm int}=\dfrac{1}{2\Sigma}(H_r+H_{\theta}).
\end{eqnarray}
Hence, by ignoring the $\mathcal{O}(\epsilon^2)$ term in \eq{eq:H0or}, there is essentially a constant term added to the unperturbed Hamiltonian $H_0$, and, therefore, the overall integrability is retained. This is given by the Hamiltonian $H_{\rm int}$. Below we present the expression for the Carter-like constant for future references. By using \eq{eq:Carter_expression} we get
\begin{equation}
\mathcal{K}=\dfrac{1}{\Sigma}\Big(r^2 H_{\theta}-a^2 \cos^2\theta H_{r}\Big).    \label{eq:CarterD}
\end{equation}
We have verified that the Poisson bracket between $\mathcal{K}$ and $H_{\rm int}$ identically vanishes, i.e., $\{\mathcal{K},H_{\rm int}\}=0$, while $\{\mathcal{K},H\}=\mathcal{O}(\epsilon^2)$, as argued before. This serves as a sanity check for the derivation of the Carter-like constant $\mathcal{K}$ in \eq{eq:CarterD}. Note that originally, the quantity $\mathcal{K}-(L_{\rm z}-aE)^2$ was described as the Carter's constant in Kerr spacetime \cite{carter68}. However, we will confine ourselves with the  definition~\eqref{eq:CarterD} throughout the rest of the paper. In addition, we will use the notation $\tilde{E}=E/m$, $\tilde{L}_{\rm z}=L_{\rm z}/m$, $\tilde{H}=H/m^2$ and $\tilde{\mathcal{K}}=\mathcal{K}/m^2$, to denote specific energy, momentum, Hamiltonian and Carter-like constant respectively. Moreover, we define $\tilde{\epsilon}=\epsilon/m=\tilde{q}B_0$. Finally, we should stress that the linearization of the Hamiltonian is only to introduce the concept of Carter-like constant in the present section. For future references in the paper, we will always use the full Hamiltonian $H$ and not $H_{\rm int}$.

Recall that unlike gravity, electromagnetism has both repulsive and attractive nature. In our system, assuming $a >0$ and $q>0$, the repulsion occurs when the external magnetic field is parallel to $L_{\rm z}$ component of the angular momentum of the body while the force is attractive when they are antiparallel. We discuss both cases until Sec.~\ref{sec:ResCros}, however, then we restrict ourselves on the case of attraction since we study this system as an analogue of an EMRI.

\section{Resonance growth} \label{sec:ResGrow}

In the previous section, we have shown that a system of non-radiating charged particle in a magnetized Kerr background preserves its integrability if the terms $\mathcal{O}(\epsilon^2)$ are neglected. Nevertheless, in the rest of paper we consider the full Hamiltonian system~\eqref{eq:hamiltonian} and we focus on particular effects, which the non-integrable perturbation has upon resonances. This section, in particular, discusses how a resonance grows as the perturbation increases, which is achieved by measuring the resonance width $w$.

The width $w$ of a resonance is formally defined as the difference between the maximum and the minimum values of the action on the separatrix of a pendulum to which a resonance can be approximated to by a normal form (for more details see \cite{Morbidelli02,LG21}). Here we adapt an approach applied in \cite{LG21} and we measure this width from rotation curves obtained by the analysis of Poincar\'{e} sections. The rest of this section outlines the employed procedure.

\subsubsection{The resonance in Poincar\'{e} section and rotation curve}
A resonance occurs when two or more characteristic frequencies of the system are commensurate. In our work, we are interested in resonances between the polar $\omega_\theta$ frequency and the radial frequency $\omega_r$. Resonances play an important role when an integrable system, like the geodesic motion on a Kerr background, is perturbed in such way that integrability is lost. According to the KAM theorem \cite{Arnold63}, the tori of the integrable system that are sufficiently away from a resonance survive the perturbation and are just slightly deformed. On such tori we observe quasiperiodic orbits which are characterized by irrational frequency ratios. On the other hand, the resonant tori in the perturbed system dissolve and only an even number of periodic orbits survive according to the {\em Poincar\'{e}-Birkhoff} theorem \cite{Birkhoff13}; half of these orbits are stable with secondary islands of stability forming around them and the other half unstable with corresponding asymptotic manifolds stemming from them. These orbits form the so called Birkhoff chain. 

\begin{figure}[htp]
    \centering
     \includegraphics[scale=.78]{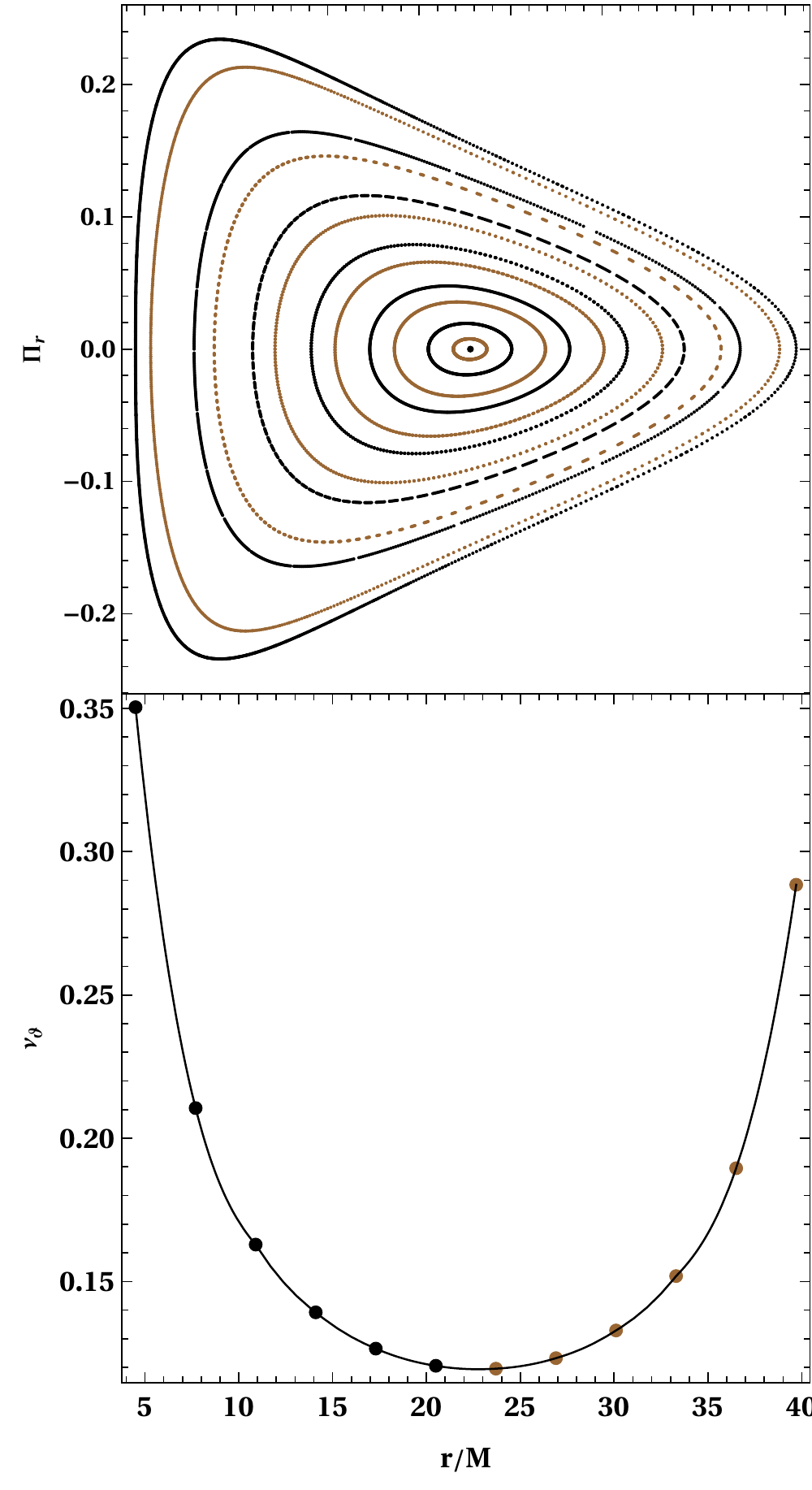}
    \caption{The top panel shows a Poincar\'{e} section for the parameters: $\tilde{E}=0.98$, $\tilde{L}_{\rm z}=3.3M$, and $\tilde{\epsilon} =-10^{-3} M^{-1}$, while the bottom panel shows the respective rotation curve when we take initial conditions along the $\pi_r=0$ line of the Poincar\'{e} section. We show $12$ initial conditions along this line starting from $r_i=4.5M$ and changing the initial distance with a step of $3.2M$. Each point on the rotation curve indicates the respective initial condition for a KAM curve depicted on the Poincar\'{e} section. The coloring indicates whether an initial condition lies left (black) or right (brown) from the center of the main island of stability.}
    \label{fig:globalPoiRot}
\end{figure}

\begin{figure*}[htp]
    \centering
    \includegraphics[scale=.8]{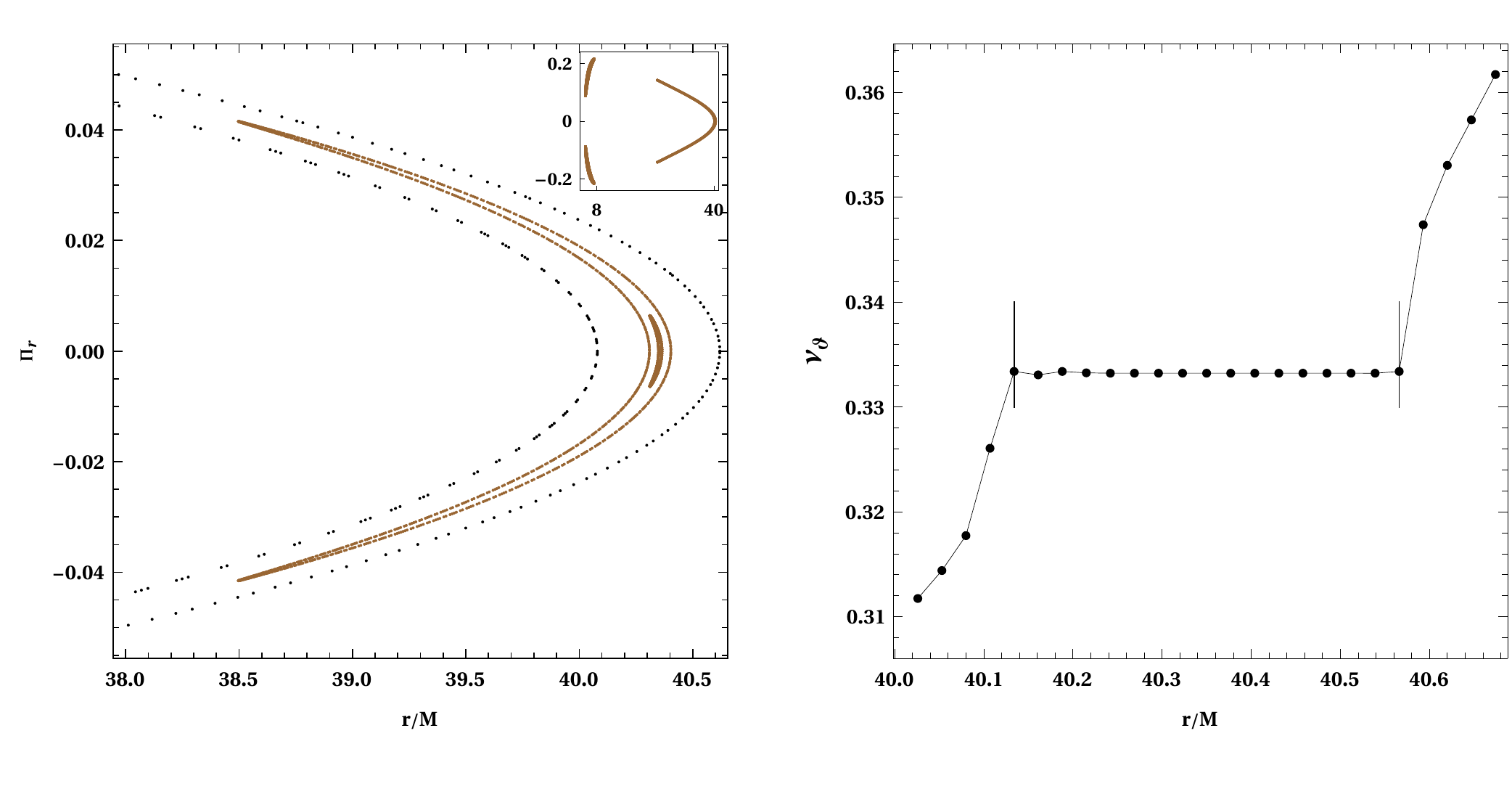}
    \caption{The left panel shows a detail from the Poincar\'{e} section shown in Fig.~\ref{fig:globalPoiRot} focusing on an island of stability of the $1:3$ resonance, while the right panel shows the respective rotation curve with the characteristic plateau. The inset plot in the left panel represents the Poincar\'{e} section for a particular initial condition. For the radial distance, we choose from $40.026M$ to $40.674M$ with a step size of $0.027M$. Note that in the right curve for the rotation number, we have included all these data points. However, for the illustration of the resonance in the left panel, we have only shown a subset of the entire data set. This is only to make the plots more reader friendly. To indicate the resonance width $w$, we have drawn two lines parallel to $\nu_\vartheta$-axis at the endpoints of the plateau. The distance between these two endpoints along the $r$-axis corresponds to the width of the resonance $w$.}
    \label{fig:resonance_1:3_B_N}
\end{figure*}

A way to investigate the resonance growth is to track them on Poincar\'{e} sections for different values of $\epsilon$, while keeping fixed the energy $E$, the angular momentum $L_{\rm z}$ and the Kerr parameter $a$. Practically, in our system a Poincar\'{e} section is formed, when we register the momentum $\pi_r$ and the radial coordinate $r$ in the Hamiltonian flow crossings of the equatorial plane with a specific orientation, e.g. when an orbit crosses the plane with $\pi_\theta>0$. Spotting a resonance on a  Poincar\'{e} section just by inspecting it is often quite difficult, see, e.g., the top panel of Fig.~\ref{fig:globalPoiRot}. To achieve it in a two degrees of freedom system, a useful tool is the rotation number \cite{Voglis98,Contopoulos02}, which provides the ratio of the characteristic frequencies.

A rotation number can be calculated from a Poincar\'{e} section. The first step is to identify a fixed point $\vec{x}_c$ on the Poincar\'{e} section, around which closed curves are nested. This point is often called the centre of the main island of stability and the curves correspond to cuts through tori, for which the frequency ratio is an irrational number. In the top panel of Fig.~\ref{fig:globalPoiRot} the position of the fixed point is marked by a black dot. In  the next step, rotation angles between successive intersections $\vec{x}_i$ with the section with respect to $\vec{x}_c$ are calculated as
\begin{equation}
    \vartheta_i = \mathrm{ang}\left[\left(\vec{x}_{i+1}-\vec{x}_c\right), \left(\vec{x}_{i}-\vec{x}_c\right)\right] \:.
\end{equation}
The rotation number is then defined as
\begin{equation}\label{eq:angular_moment}
    \nu_\vartheta = \lim_{n\to\infty} \frac{1}{2\pi n}\sum_{i=1}^n \vartheta_i \:.
\end{equation}
For finite $n$ the accuracy of the rotation number is of the order of $1/n$. For an integrable non-degenerate Hamiltonian system, the rotation number changes monotonically for initial conditions getting radially further away from $\vec{x}_c$. The respective curve is called rotation curve. When the integrability is broken, then at the resonances the rotation curve fluctuates randomly, when it is calculated on a chaotic layer, or provides a plateau, when it is calculated on a secondary island of stability. The bottom panel of Fig.~\ref{fig:globalPoiRot} shows how the rotation curve looks like when we scan the section depicted in the top panel along $\pi_r=0$.

In both panels of Fig.~\ref{fig:globalPoiRot}, we do not see direct signs of a resonance. However, the rotation curve indicates where we have to look to find one. For example, to find the $1:3$ resonance, one has to look between the first two initial conditions from the left of Fig.~\ref{fig:globalPoiRot} or beyond the first initial condition from the right. Doing the latter provides a detail of the Poincar\'{e} section shown in the left panel of Fig.~\ref{fig:resonance_1:3_B_N}. The inset of this panel provides all the three expected islands, while the main panel focuses just on one of the three islands of stability. The respective rotation curve is depicted in the right panel of  Fig.~\ref{fig:resonance_1:3_B_N}, in which one can spot easily the characteristic plateau of the resonance. The length of this plateau measured along the radial coordinate $r$ provides an adequately good measurement of the width of the resonance $w$.

\subsubsection{Growth of resonance}

\begin{figure}[htp]
    \centering
    \includegraphics[width=0.95\linewidth]{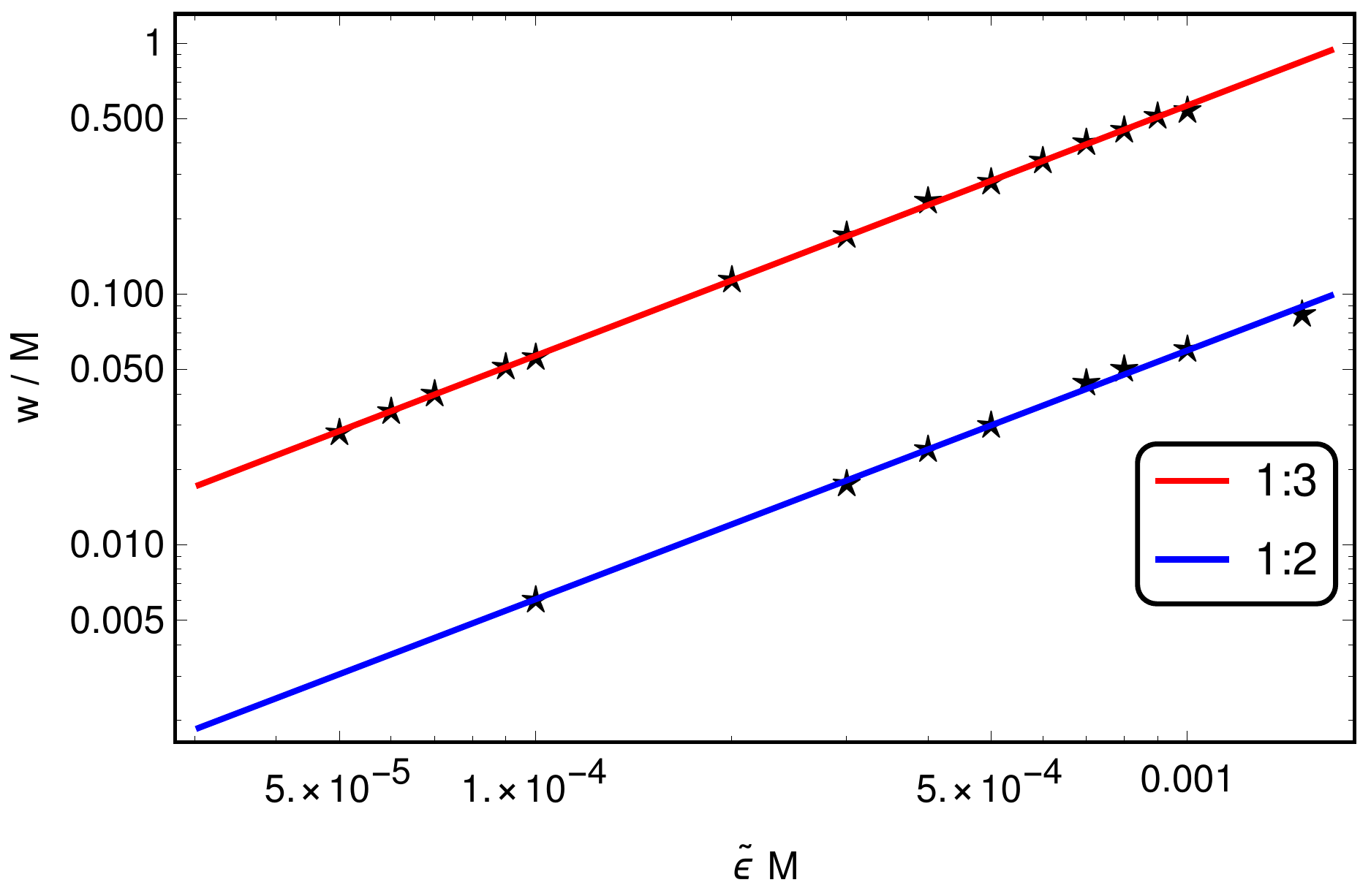}
    \caption{The growth of the width of resonances as a function of the parameter $\epsilon$ in a log-log plot. The stars indicate the actual cases studied, while the lines are interpolations through these cases. To construct the above figures, we have used the following set of parameters: $\tilde{E}=0.98$, $\tilde{L}_{z}=3.8M$, and $a=0.5M$.}
    \label{fig:resonance_width}
\end{figure}

As was already mentioned, resonances in dynamical systems can be approximated locally by the dynamics of a pendulum \cite{Morbidelli02}. Using the above fact one can find that the square of the width of a resonance is proportional to the perturbation parameter \cite{Morbidelli02,Zelenka20,LG21}. By plotting the width of a resonance as a function of $\epsilon$ as we do in Fig.~\ref{fig:resonance_width}, we can correlate the perturbation parameter with $\epsilon$. We investigate two main resonances ($1:2$ and $2:3$) and the width of the resonances for each $\epsilon$ is determined by the length of each plateau on the corresponding rotation curve \cite{LG21}.

In Fig.~\ref{fig:resonance_width} we fitted the data points for each resonance with the curve,
\begin{equation}
    \log(w)=A+B \log(\epsilon).
\end{equation}
For the two under study $1:3$ and $1:2$ resonance, we found that $B=0.997577$, and $B=0.992058$ respectively. Given the slope in both cases is $1$ (within a numerical error of $1\%$), we deduce that the perturbation parameter driving the system away from integrability is proportional to $\epsilon^2$. In other words, the system is integrable up to $\mathcal{O}(\epsilon)$. 

This confirms our finding in Sec.~\ref{sec:ConQuant}, where we obtained that the Carter-like constant is valid up to $\mathcal{O}(\epsilon)$. It is interesting to note that in this feature the system is similar to the case of a spinning body moving in a Kerr black hole background. Namely, there is a Carter-like constant valid up to linear order in the spin of the secondary \cite{rudiger1981conserved,rudiger1983conserved,gibbons1993susy,Tanaka:1996ht,Witzany19}, while the perturbation parameter driving the system to non-integrability appears to be proportional to the square of the spin \cite{Zelenka20}.

\section{Effects of the electromagnetic self-force}\label{sec:EMsf}

The motion of accelerated charged body constitutes an interesting theoretical problem and has a long-standing history \citep{rohrlich2020classical}. Starting with the seminal works of Lorentz, Abraham and Poincar{\'e} in the Newtonian case \cite{mcdonald2018history}, the contributions from Dirac \citep{Dirac:1938nz}, Landau \citep{landau1975classical}, Dewitt, and Brehme for the relativistic domain have made remarkable expansion of the field \cite{dewitt1960radiation}. Recent times have also witnessed a significant growth of interest in this topic \cite{kolos21,tursunov18}. For an excellent review we refer to \citep{poisson04}. In this section, we introduce the respective equations of motion in a ready to use format, and we discuss the resonant crossing effects for different initial conditions.

\subsection{Equations of motion with the self-force} 

Before delving into the relativistic case, let us first discuss the Newtonian counterpart. In this limit, the equations of motion are given by Abraham-Lorentz equation \citep{lorentz1892theorie}
\begin{equation}
    m\dfrac{d\vec{\mathcal{V}}}{dt}=\vec{F}_{\rm ext}+\dfrac{2q^2}{3}\dfrac{d^2\vec{\mathcal{V}}}{dt^2},
\end{equation}
where $\vec{\mathcal{V}}$ is the velocity, $\vec{F}_{\rm ext}$ is the external Lorentz force, and $t$ is the time. Note that the second term on the right side captures the effect of the self-force of the moving charge, and it is provided by a derivative one order higher than the left side. This would lead to runaway solutions, which are physically inconsistent. One easy way to appreciate this is to switch off the external force, i.e.,  $F_{\rm ext}=0$, and we obtain $\mathcal{V} \sim \exp[3mt/(2q^2)]$. This diverges as $t$ approaches infinity, and leads to a unphysical system. In order to avoid this pathology, one typically adopt the approach introduced by Landau and Lifshitz in \citep{landau1975classical}, that is known as an \enquote{order reduced} formalism. Within this approximation, the above expression can be written as:
\begin{equation}
       m \dfrac{d\vec{\mathcal{V}}}{dt}=\vec{F}_{\rm ext}+\dfrac{2q^2}{3m}\dfrac{d \vec{F}_{\rm ext}}{dt}.
\end{equation}
In the case of Lorentz force, we have
\begin{align}
  \vec{F}_{\rm ext}=q(\vec{E}_{\rm ext}+\vec{\mathcal{V}} \times \vec{B}_{\rm ext}),  
\end{align}
where $\vec{E}_{\rm ext}$ and $\vec{B}_{\rm ext}$ are the electric and magnetic field respectively. If we assume that these fields are independent of time, we arrive at 
\begin{eqnarray}
 \dfrac{d\vec{\mathcal{V}}}{dt}&=&\vec{F}_{\rm ext}+\dfrac{2q^3}{3m}\Big(\dfrac{d\vec{\mathcal{V}}}{dt} \times \vec{B}_{\rm ext}\Big), \nonumber \\
 &=& \vec{F}_{\rm ext}+\dfrac{2q^3}{3m^2}\Big(\vec{F}_{\rm ext} \times \vec{B}_{\rm ext}\Big).
\end{eqnarray}
Once the external fields are known, we can obtain the final trajectory of the body. Therefore, this approach gives a self-consistent way to deal with the electromagnetic self-force of a charged body.

The relativistic correction to the Abraham-Lorentz equation was first introduced by Dirac and it is known as Abraham-Lorentz-Dirac equation \citep{Dirac:1938nz}
\begin{equation}
     m\dfrac{d\mathcal{U}^{\mu}}{d\tau}= F^{\mu}_{\rm ext}+ \dfrac{2q^2}{3}\Big(\delta^{\mu}_{\nu}+\mathcal{U}^{\mu}\mathcal{U}_{\nu}\Big)\dfrac{d a^{\nu}}{d\tau},
     \label{eq:ESF_Flat}
\end{equation}
where, $F^{\mu}_{\rm ext}$ is the external Lorentz force, and given as $ F^{\mu}_{\rm ext}=qF^{\mu}_{~\nu}~\mathcal{U}^{\nu}$, and $a^{\mu}$ is the acceleration vector. By following the \enquote{order reduced} approach discussed in the Newtonian case, we can write
\begin{equation}
    a^{\mu}=(1/m)F^{\mu}_{\rm ext}=\dfrac{q}{m}F^{\mu}_{~\nu}\mathcal{U}^{\nu}.
\end{equation}
With the above substitution, we obtain
\begin{eqnarray}
 \dfrac{da^{\mu}}{d\tau}= \dfrac{d^2\mathcal{U}^{\mu}}{d\tau^2}&=&\dfrac{q}{m}\Big\{\Big(\dfrac{dF^{\mu}_{~\nu}}{d\tau}\Big)\mathcal{U}^{\nu}+(q/m)F^{\mu}_{~\nu}F^{\nu}_{~\alpha}\mathcal{U}^{\alpha}\Big\}, \nonumber \\
  &=& \tilde{q}\dfrac{dF^{\alpha}_{~\beta}}{dx^{\mu}}\mathcal{U}^{\mu}\mathcal{U}^{\beta}+\tilde{q}^2 F^{\alpha}_{~\beta}F^{\beta}_{~\mu}\mathcal{U}^{\mu}. \label{eq:Double_U_der}
\end{eqnarray}
Finally, we can rewrite the above expression in a more reader friendly way as follows:
\begin{widetext}
\begin{equation}
 \dfrac{d\mathcal{U}^{\mu}}{d\tau}=\tilde{q}F^{\mu}_{~\nu}\mathcal{U}^{\nu}+\dfrac{2{q}^2}{3m}\Big(\delta^{\mu}_{\nu}+\mathcal{U}^{\mu}\mathcal{U}_{\nu}\Big)  \Big(\tilde{q}\dfrac{dF^{\nu}_{~\beta}}{dx^{\gamma}}\mathcal{U}^{\gamma}\mathcal{U}^{\beta}+\tilde{q}^2 F^{\nu}_{~\beta}F^{\beta}_{~\gamma}\mathcal{U}^{\gamma}\Big). 
\end{equation}
\end{widetext}
For our future reference, we will call the prefactor of the second term in Eq. (\ref{eq:ESF_Flat}), i.e., $2q^2/3m$ as \textit{radiation} parameter $k$, which actually scales the electromagnetic self-force felt by the charged body. In the case of EMRI driven by GSF, the analogous parameter would be the mass ratio $\eta=m/M$. In our system, the both parameters are formally related as follows:
\begin{equation}
    \dfrac{k}{M}=\dfrac{2}{3}\tilde{q}^2\eta,
\label{radiation_parameter}
\end{equation}
Note that even if we fix the value of $k$ (with respect to $M$), the mass ratio still depends on the arbitrary choice of specific charge. For instance, in our subsequent numerical examples we employ the value $k=10^{-3}M$, for which $\eta \sim 10^{-1}(\tilde{q})^{-2}$. However, one has to keep in mind that the GSF and the electromagnetic self-force (ESF) are intrinsically different. Therefore, the mass ratio found from the above formula is not directly relevant for GSF applied in EMRI systems.

In the curved spacetime, the self-forced motion of charged body is derived by DeWitt and Brehme \citep{dewitt1960radiation}:
\begin{equation}
    \dfrac{D\mathcal{U}^{\mu}}{d\tau}=\tilde{q}F^{\mu}_{~\nu}~\mathcal{U}^{\nu}+f^{\mu}_{R},
    \label{eq:Lorentz_Dirac}
\end{equation}
where $f^{\mu}_{R}$ is the radiation reaction given by the following expression:
\begin{eqnarray}
f^{\mu}_{R}&=& k\Big(\dfrac{D^2\mathcal{U}^{\mu}}{d\tau^2}+\mathcal{U}^{\mu}\mathcal{U}_{\nu}\dfrac{D^2\mathcal{U}^{\nu}}{d\tau^2}\Big)+\nonumber \\
&& \dfrac{q^2}{3m}\Big(R^{\mu \nu}\mathcal{U}^{\nu}+R^{\nu}_{\lambda}\mathcal{U}_{\nu}\mathcal{U}^{\lambda}\mathcal{U}^{\mu}\Big)+3k f^{\mu \nu}_{\rm tail}\mathcal{U}_{\nu}, \label{eq:fmuR}
\end{eqnarray}
with $f^{\mu \nu}_{\rm tail}$ being the tail term:
\begin{equation}
    f^{\mu \nu}_{\rm tail}=\int^{\tau-0^{+}}_{-\infty} D^{[\mu}G^{\nu ]}_{+ \lambda^{\prime}}(z(\tau),z(\tau^{\prime}))\mathcal{U}^{\lambda^{\prime}}d\tau^{\prime}. \label{eq:tail}
\end{equation}
The above expression contains an integral over the entire past of the body. 

In the present case, we can simplify Eq.~\eqref{eq:Lorentz_Dirac} further by considering the advantage of working in vacuum, and set the Ricci tensor to zero. Therefore, the second parenthesis in \eq{eq:fmuR} vanishes and does not contribute. Moreover, by following the discussion in \cite{kolos21} and \cite{tursunov18}, we can also neglect the tail term, and finally arrive at:
\begin{equation}
    \dfrac{D\mathcal{U}^{\mu}}{d\tau}=
    \tilde{q}F^{\mu}_{~\nu}~\mathcal{U}^{\nu}+k\Big(\dfrac{D^2\mathcal{U}^{\mu}}{d\tau^2}+\mathcal{U}^{\mu}\mathcal{U}_{\nu}\dfrac{D^2\mathcal{U}^{\nu}}{d\tau^2}\Big).
    \label{eq:ESF_Charged}
\end{equation}
We note that in its gravitational analogue (i.e., in EMRI driven by GSF), the tail term (which corresponds to the effect of the backscattered gravitational radiation) remains relevant. Nevertheless, in the present electromagnetic model, its contribution is negligible as further discussed in Sec.~\ref{sec:Param}, hence, we ignore this term in our calculations. This approximation is sufficient for our study as the key objective of our work is to study resonance crossing in a EMRI analogue model with the help of a slow dissipation introduced by ESF.

Note that Eq.~(\ref{eq:ESF_Charged}) matches with the flat spacetime relation given in Eq. (\ref{eq:ESF_Flat}), except for the fact that the ordinary derivative is now replaced with the covariant derivative. With this in mind, the expression for $D^2\mathcal{U}^{\nu}/d\tau^2$ can also be obtained directly from Eq.(\ref{eq:Double_U_der}) by switching to covariant derivatives: 
\begin{equation}
    \dfrac{D^2\mathcal{U}^{\mu}}{d\tau^2}=\tilde{q}\dfrac{DF^{\mu}_{~\nu}}{dx^{\gamma}}\mathcal{U}^{\gamma}\mathcal{U}^{\nu}+\tilde{q}^2 F^{\mu}_{~\nu}F^{\nu}_{~\gamma}\mathcal{U}^{\gamma}.
    \label{eq:ESF_Charged2}
\end{equation}
Unlike the Hamiltonian  system discussed in Sec.~\ref{sec:ChargedPartIntro} where the single parameter $\tilde{\epsilon}$ is sufficient to describe the interaction with the EM field, here the situation becomes slightly more complicated and two independent parameters are needed. Namely, we have to deal with $\tilde{\epsilon}$ and $k$,  when the radiation reaction is taken into account. Indeed, by inspecting Eqs.~\eqref{eq:ESF_Charged} and \eqref{eq:ESF_Charged2}, we observe that although $B_0$ again couples with $\tilde{q}$ in both terms, the multiplication by the factor $k$ is present only in the second term. Therefore, we need to set both values $\tilde{\epsilon}$ and $k$ independently.

Moreover, by inserting \eqref{eq:ESF_Charged2} into \eqref{eq:ESF_Charged}, one notices that the leading perturbation parameter is $\epsilon$, which comes from the Lorentz force Eq.~\eqref{eq:LORENTZ} and the self-force introduces higher order perturbations, i.e. $k \epsilon$ and $k \epsilon^2$. This indicates that even if the self-force is taken into account, the width of the resonance is mainly defined by $\epsilon$.

In all the future references to ESF, we consider the equation of motion given by \eq{eq:ESF_Charged} and we parametrize the dissipating trajectories by $\tilde{\epsilon}$,$k$ and the initial values of $\tilde{E}$ and $\tilde{L}_z$.

\subsection{Resonance crossings} \label{sec:ResCros}

\begin{figure}[htp]
\vspace{0.5cm}
\includegraphics[width=0.95\linewidth]{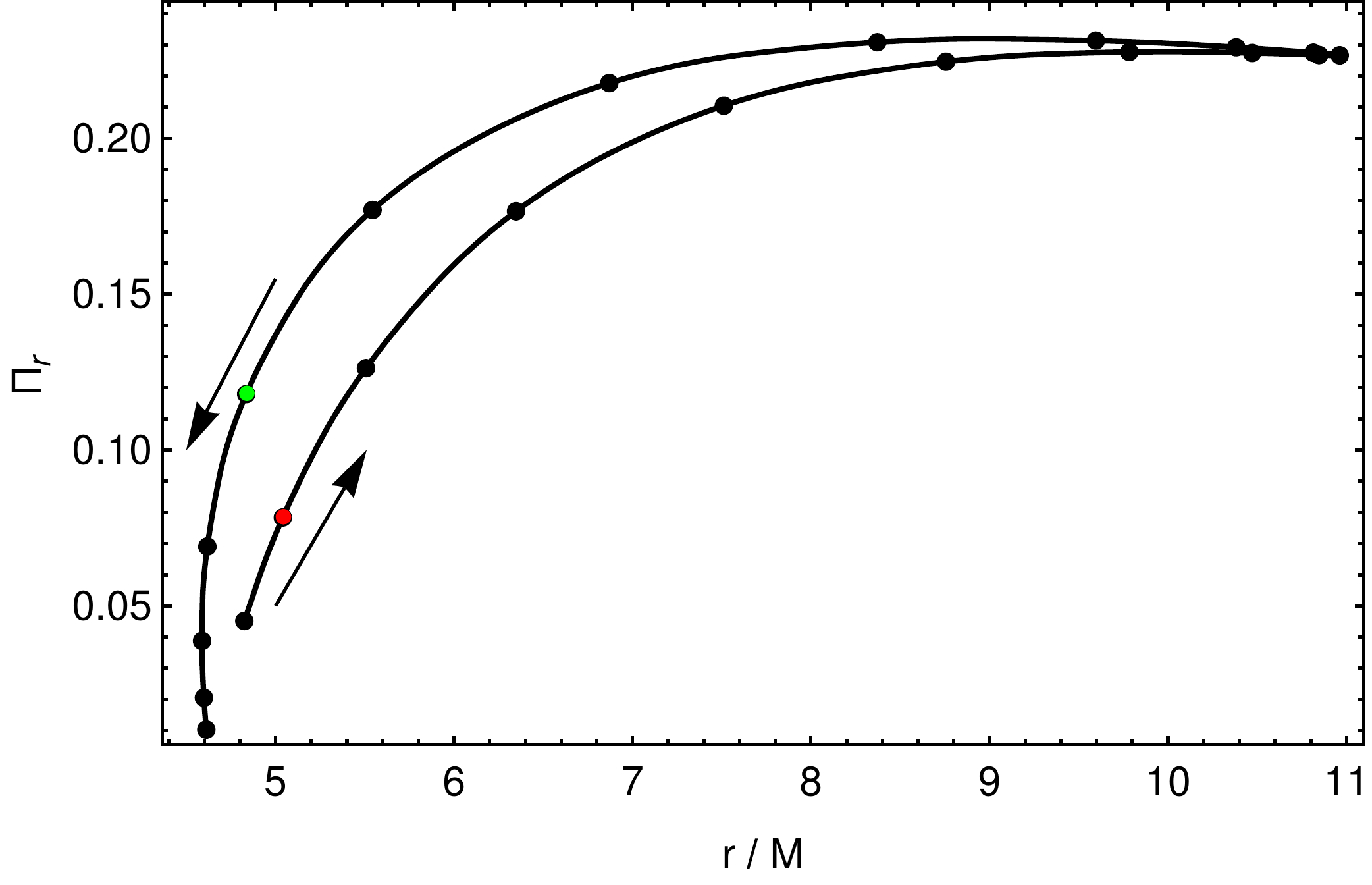}
\includegraphics[width=0.95\linewidth]{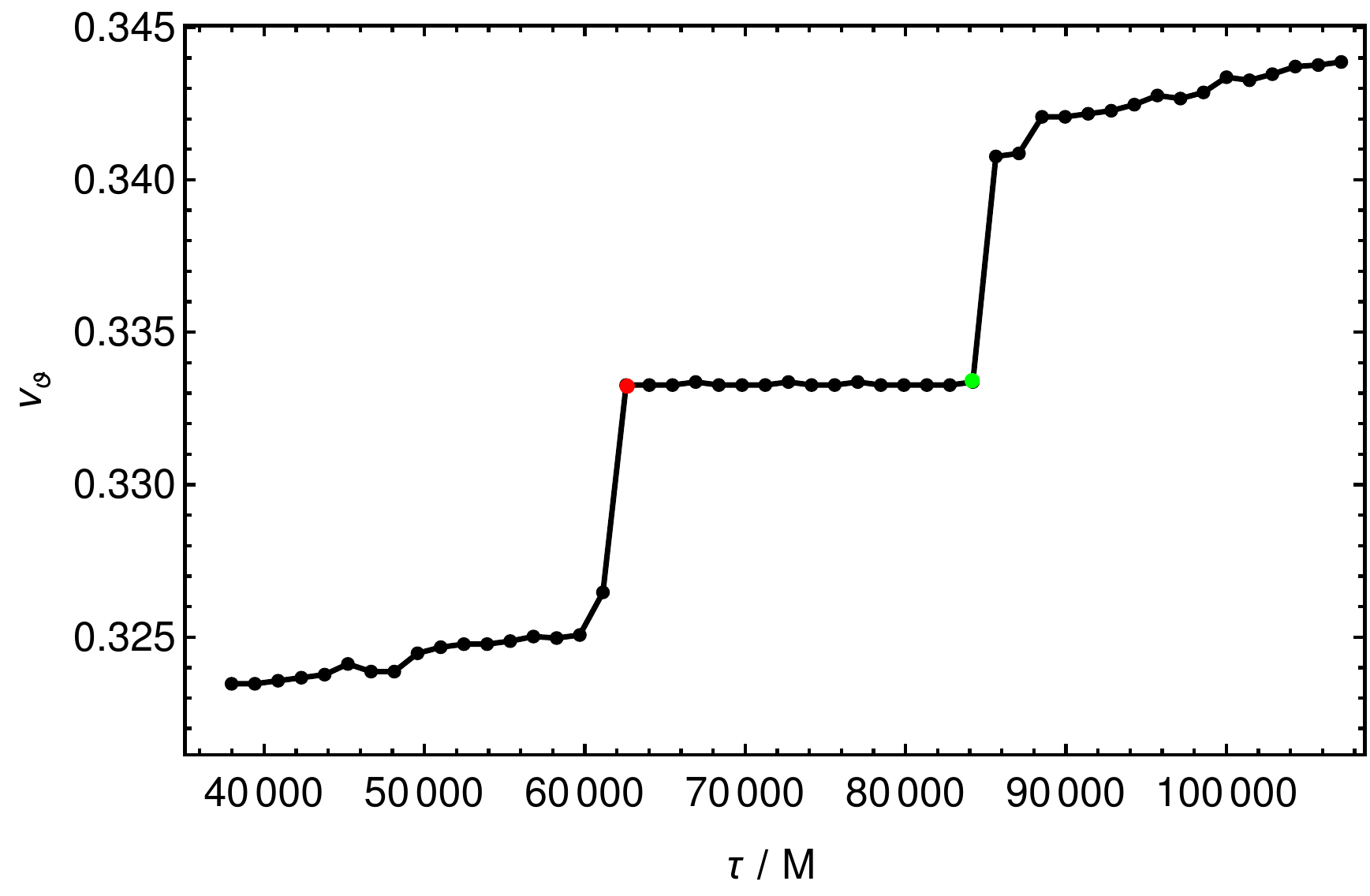}
\caption{The plots show a case of a resonance crossing in the dissipative system with an attractive Lorentz force. The trajectory of the inspiral starts with $\tilde{E}=0.98$, $\tilde{L}_{\rm z}=3.3M$, $\pi_r=0$, $\theta=\pi/2$, initial value of the radial coordinate is chosen to be $r=40.0995M$, and the rest parameters of the system are $\tilde{\epsilon}=-10^{-3} M^{-1} $, and $k= 10^{-3} M$. The top panel shows a stroboscopic depiction of the Poincar\'{e} section, with the red dot indicating the entrance of the inspiral into the resonance and the green the exit. The arrows indicate the way the stroboscopic depiction evolves on the section. The bottom panel shows the rotation curve, i.e. the rotation number as a function of the proper time. The $1:3$ resonance is indicated by the presence of the characteristic plateau on the rotation curve.}
\label{fig:KBH_attractive_dissip}
\end{figure}

\begin{figure*}[htp]\centering
\includegraphics[width=0.45\textwidth]{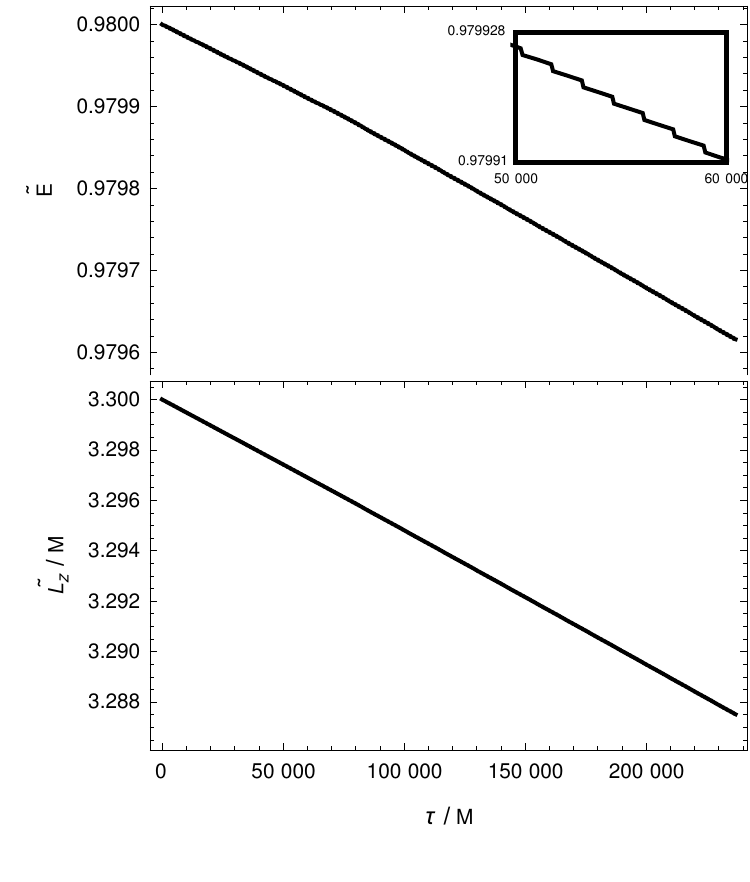}
\includegraphics[width=0.45\textwidth]{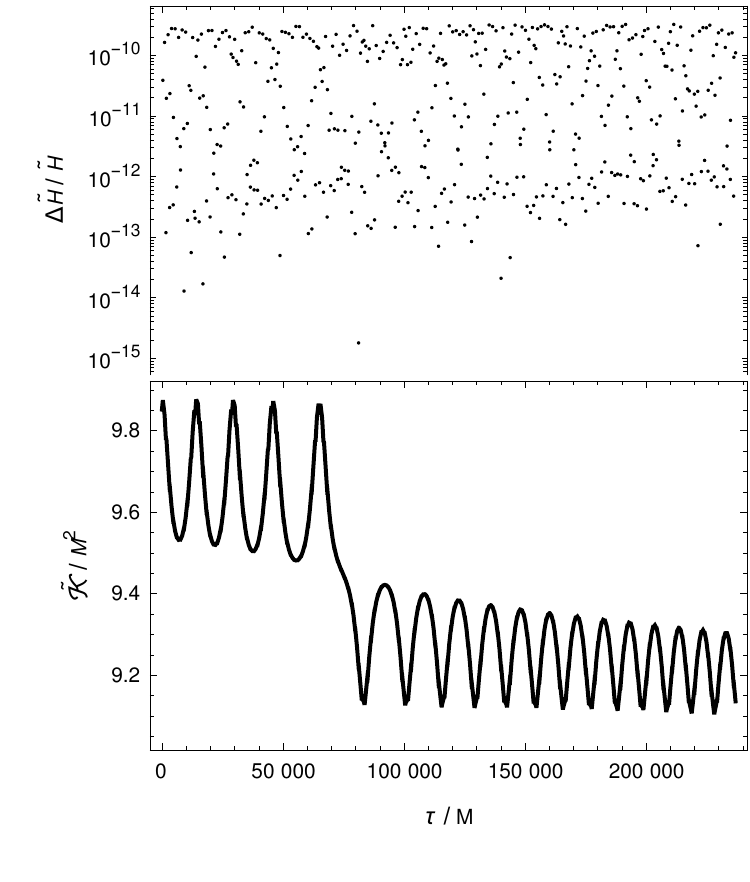}
\caption{Dissipation for various adiabatically changing quantities are shown: the initial values of different parameters are: $r_i=40.09950~M$, $\tilde{E}=0.98$, and $\tilde{L_{\rm z}}=3.3M$. The top left shows the dissipation of the energy with the inset focusing on a small part of the graph so that the oscillations become visible; the bottom left  shows how the angular momentum is dissipating over time; the top right panel shows that the variation of the relative error $\Delta\tilde{H}/\tilde{H}$ is $\sim \mathcal{O}(10^{-15})$ to $\mathcal{O}(10^{-10})$. The bottom right depicts the evolution of $\tilde{\mathcal{K}}$, as given in Eq.~\eqref{eq:Carter_expression}. This evolution differs significantly from the other quantities and it} is easy to spot the jump which corresponds to the resonance.
\label{fig:DissConst}
\end{figure*}

From the Hamiltonian system, we are already informed about the  existence of resonant islands for different initial conditions. We can place a body on an initial condition right outside a known resonance layer and dissipate the system using the self-force~\eqref{eq:ESF_Charged}. The expectation is that the inspiralling body at a point will hit the resonance and cross it. Such a resonance crossing is shown in Fig.~\ref{fig:KBH_attractive_dissip}. Since the Hamiltonian system is non-integrable, in accordance to the EMRI terminology we call these resonances prolonged \cite{Zelenka20,LG21}. 

The top panel of Fig.~\ref{fig:KBH_attractive_dissip} shows a stroboscopic depiction of a $1:3$ resonance crossing on a Poincar\'{e} section\footnote{Note that since the system is dissipating the use of the term Poincar\'{e} section is approximative, i.e. a loan from the Hamiltonian non-dissipative case.}, which means that we used only every third point in the section's sequence. Using each point of this sequence as an initial condition we can evolve the system without the self-force in order to find the rotation number for each of these points as was also done in \cite{LG21}. The result is a rotation curve shown in the bottom panel of  Fig.~\ref{fig:KBH_attractive_dissip}, where the rotation numbers are plotted with the respect to the proper time.  The plateau at $1:3$ indicates the points corresponding to the crossing of the resonance.

With the dissipation turned on, previously constant quantities will be evolving. Fig.~\ref{fig:DissConst} shows a typical evolution of the energy $\tilde{E}$, the angular momentum $\tilde{L}_{\rm z}$, the relative error of the Hamiltonian $\Delta\tilde{H}/\tilde{H}$ and $\tilde{\mathcal{K}}$ \footnote{Recall that $\tilde{\mathcal{K}}$ is actually not a constant when the system is non-linear in $\epsilon$.}. The energy and the angular momentum follow an almost linear decline, while the Hamiltonian is conserved up to numerical precision. The behavior of the quantity $\tilde{\mathcal{K}}$ deserves more attention. Unlike the other three quantities, it is actually not an integral of motion and its value oscillates even without the self-force. Moreover, if the self-force is introduced, $\tilde{\mathcal{K}}$ exhibits an abrupt drop during the resonance crossing. Such jumps are quite common at resonance crossings induced by GSF \cite{Flanagan12,Berry16} and they typically scale as the square root of the parameter which perturbs the system and induces dissipation (i.e., the square root of the mass ratio in the case of GSF). In our model, the perturbation parameter $\tilde{\epsilon}$ is independent of the self-force effects captured by the parameter $k$. As was already discussed in Sec.~\ref{sec:EMsf}, the width of the resonance should be determined mainly by $\tilde{\epsilon}$. A reasonable expectation is that the jump in $\tilde{\mathcal{K}}$ should be proportional to the width of the resonance. To confirm this, we did some numerical checks. For resonance $1:2$, with $\tilde{\epsilon}=10^{-3}M^{-1}$, and $r_i=41.2M$, the jump in  $\tilde{\mathcal{K}}$ (the change in its value between entering and leaving the resonance) is $\sim 0.08M^2$ which is approximately $\sim \sqrt{\tilde{\epsilon}M}$. For resonance $1:3$, the jump in $\tilde{\mathcal{K}}$ for $r_i=40.098M$ and $\tilde{\epsilon}=10^{-3}M^{-1}$ is $\sim 0.6M^2$, i.e. an order of magnitude higher than $\sqrt{\tilde{\epsilon}M}$. We speculate that this discrepancy is caused by different coefficients entering the proportional relation between the width and the jump. However, this demands a meticulous investigation, which we plan for a future work. 

\section{Comparison between the adiabatic approximation and full self-force} \label{sec:adiabatic}

\begin{figure}[htp]
\includegraphics[width=0.45\textwidth]{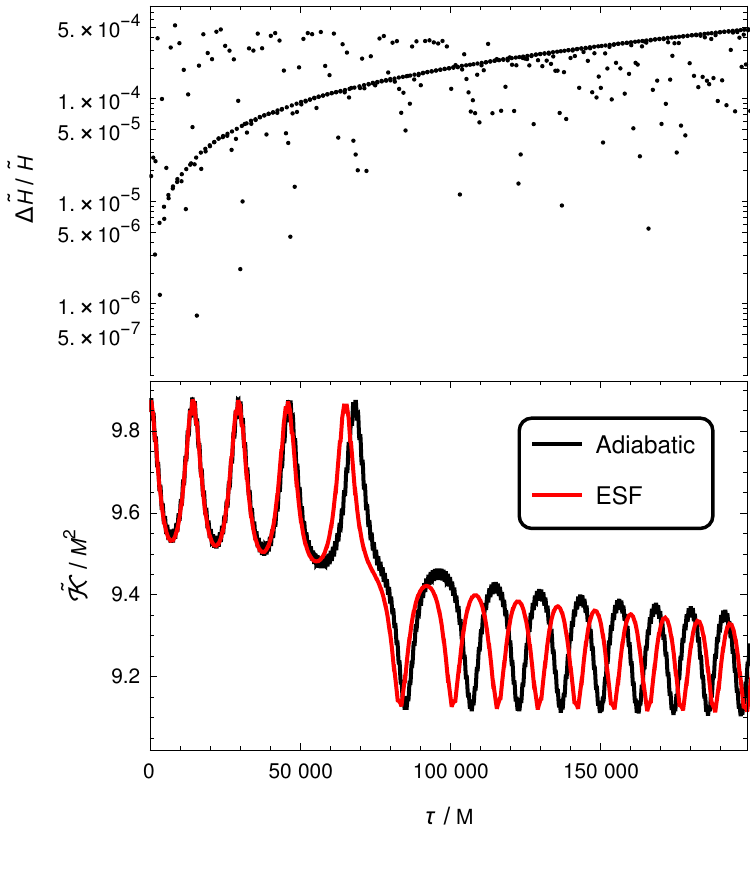}
\caption{The change in the relative error of the Hamiltonian (top panel) and of the  $\tilde{\mathcal{K}}$ quantity (bottom panel) as the binary evolves within the adiabatic approximation. The initial conditions are similar to what being used in Fig. \ref{fig:DissConst}. We use a fourth order polynomial fit for the energy and momentum, and the fitting parameters for the energy and momentum are: $a_1=-1.42 \times 10^{-9}, a_2=-1.37 \times 10^{-15}, a_3=2.17 \times 10^{-21}, b_1=-5.1 \times 10^{-8}, b_2=-1.06 \times 10^{-14}, b_3=1.38 \times 10^{-20}$.}
\label{fig:DissConstADB}
\end{figure}

 \begin{figure*}[htp]
\includegraphics[width=0.45\linewidth]{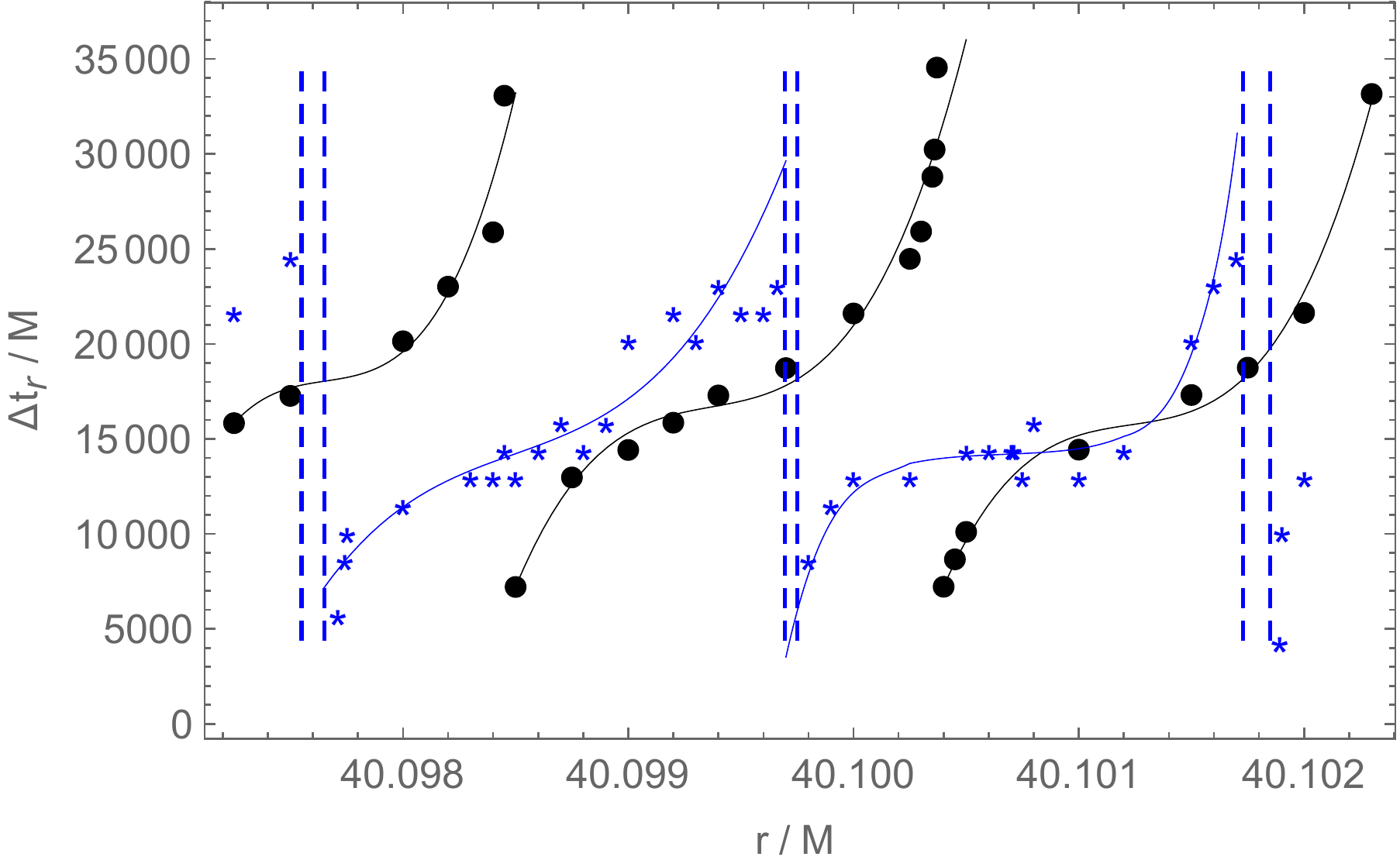}
\includegraphics[width=0.44\linewidth]{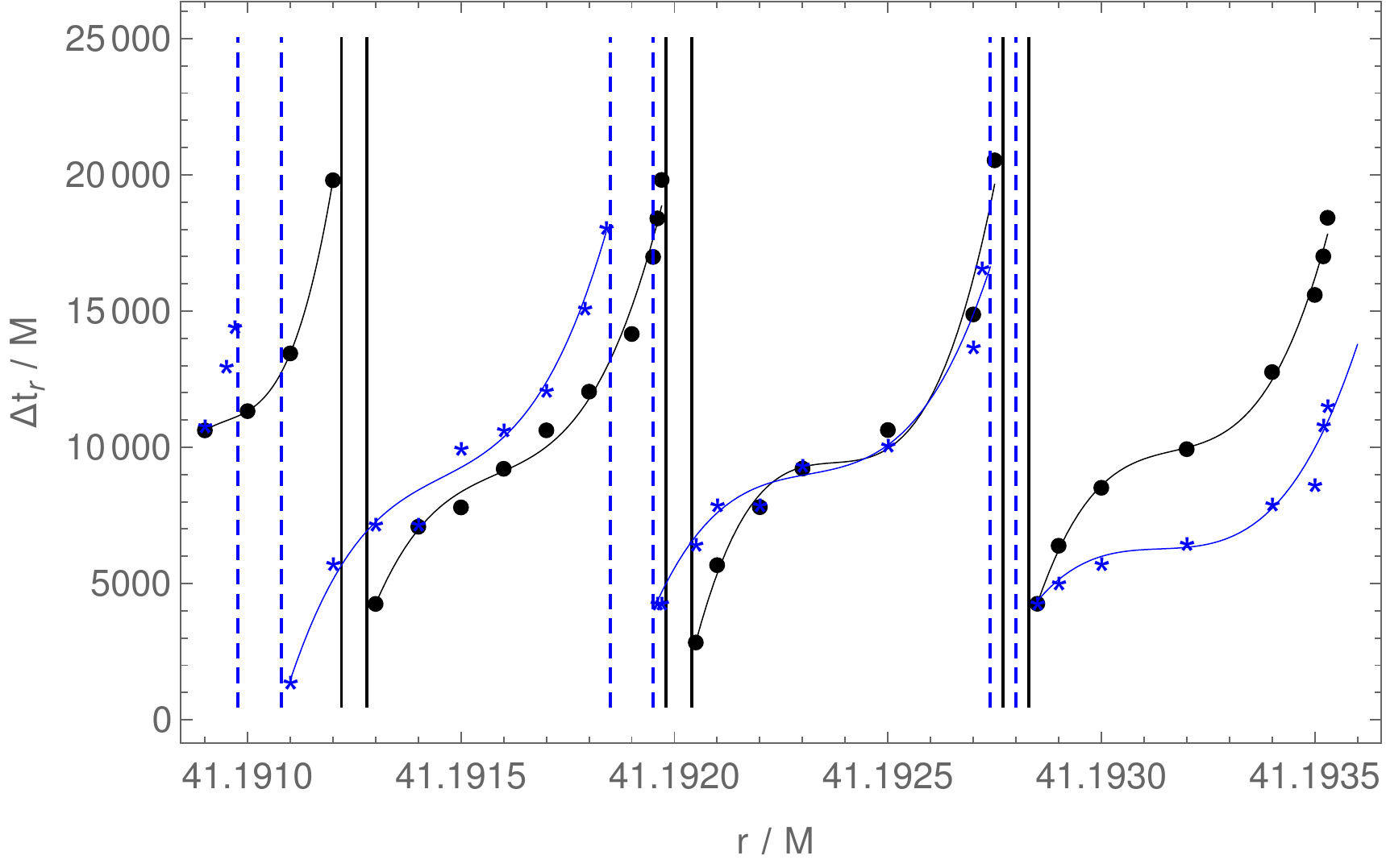}
\caption{Periods of time $\Delta t_r$ spent by a body within the resonance (as a function of the initial radius) comparing the full self-force description (black) with the approximated adiabatic evolution (blue). The space between two nearby vertical dashed blue lines indicates the initial radii for which the body following the adiabatic evolution enters the resonance and we do not see it leaving it. In the right plot the space between two nearby vertical continuous black lines indicate the same thing for the self-force driven evolution. The left plot shows the $1:3$ resonance, while the right the $1:2$. The following values for the parameters were used: $\tilde{E}(0)=0.98$, $\tilde{L_{\rm z}}(0)=3.8M$, $a=0.5M$, $\tilde{\epsilon}=-10^{-3}M^{-1}$, and $k=10^{-3}M$.}
\label{fig:ResTime}
\end{figure*}

One of the key objectives of this paper is to compare adiabatic approximation with the instantaneous electromagnetic self-force computations. This may hint us to shape our ideas in the GSF sector as well. As the application of self-force is relatively easier in the electromagnetic case than its gravitational counterpart, we can exploit this advantage.
For the adiabatic approximation, we have not followed the traditional approach to obtain the fluxes due to energy and angular momentum, as typically done in literature \cite{Drasco06}. Namely, in the traditional adiabatic approach of an EMRI the evolution of the system tracks the slow dissipation of the constants of motion: energy, momentum and Carter constant. Since the geodesic motion in a Kerr background is integrable, one can correlate the values of the constants with the orbital parameters of the body and track the inspiral. In our case, however, the system (unless being linearized in $\epsilon$) is non-integrable and lacking the Carter-like constant. To employ an adiabatic scheme for the non-integrable system and compare it with the full self-force results we do the following. We use the energy and angular momentum values obtained from the instantaneous self-force to fit the respective data sets, and obtain the energy and momentum as functions of time. In other words, we can write energy and momentum as follows:
\begin{eqnarray}
 && E(\tau)=E(0)+\sum^{N}_{n=1}a_{n}\tau^n \nonumber \\
 && L_{\rm z}(\tau)=L_{\rm z}(0)+\sum^{N}_{n=1}b_{n}\tau^n
 \label{eq:fitted_energy}
\end{eqnarray}
where $a_n$ and $b_n$ are usual expansion coefficients capturing the effects of the self-force, and $E(0)$ and $L_{\rm z}(0)$ are the initial values of energy and momentum respectively. The instantaneous self-force comes with the advantage of adding as many order of correction as we want, and it is only a matter of higher order fitting, i.e. higher $N$. In an example discussed in the Appendix, we indicate that the order of fitting may affect the final outcome of the resonance crossing. Our numerical investigation showed that the mismatch is marginal if we consider $N \geq 2$. In most cases when reproducing the plots, we consider the fitting to be a fourth order polynomial $(N=4)$. 

Our adiabatic scheme is similar with those used in \cite{LGAC10,LG21,Destounis21,Destounis21b}, the only difference is that those studies used averaged fluxes instead of using the instantaneous self-force. As in these studies the absence of a Carter-like constant limits our dissipation scheme in using just the energy and the angular momentum losses. By using the fitted function of the energy and momentum as pointed out in Eq.~\eqref{eq:fitted_energy}, we obtain $P^{t}$ and $P^{\phi}$ as a function of time:
\begin{eqnarray}
    P^{t} &=& E(\tau)+\dfrac{2Mr \Big[E(\tau)(r^2+a^2)-aL_{\rm z}(\tau)\Big]}{\Delta \Sigma}-a\epsilon, \nonumber \\
    P^{\phi} &= &\dfrac{L_{\rm z}(\tau)\csc^2\theta}{r^2+a^2}+\dfrac{2 a M r\Big[E(\tau)(r^2+a^2)-aL_{\rm z}(\tau)\Big]}{(r^2+a^2)\Delta \Sigma} \nonumber \\
    & & -\epsilon/2\, .
\end{eqnarray}
The self-force corrections are encoded within the time evolution of energy and angular momentum. To obtain the other components namely $P^r$ and $P^{\theta}$, we use Eq.~\eqref{eq:LORENTZ}, which does not include the dissipative effects. The entire premise of using the adiabatic approximation here is to averaged out the instantaneous self-force contribution and ignore some of its components. By comparing the results obtained from the adiabatic evolution with the self-force one, we intend to deduce arguments relevant for the EMRI gravitational counterpart.

In Fig.~\ref{fig:DissConstADB}, we consider a typical example where the adiabatic evolution is shown. The detailed parameter space and the fitting parameters are given in the caption of the plot. If we ignore the scattered points, we can see that the relative error in the Hamiltonian steadily grows (top panel of Fig.~\ref{fig:DissConstADB}) in contrast to the self-force calculation (right top panel of Fig.~\ref{fig:DissConst}). The latter shows that the Hamiltonian remains conserved up to $\mathcal{O}(10^{-10})$, which implies a numerical precision accuracy. Hence, one of the side-effects of using our scheme for adiabatic approximation is that the mass of the inspiralling body changes with time.

The evolution of $\tilde{\mathcal{K}}$ is also affected by the adiabatic approximation as can be seen form the bottom plot of Fig.~\ref{fig:DissConstADB}, where it is compared with the instantaneous electromagnetic self-force calculations. In particular, the difference grows in time, and becomes more prominent near the resonance. This might be an artifact of the employed adiabatic approximation, which could be enhanced by the fact that we ignored the evolution of $\tilde{\mathcal{K}}$. It has been shown by Isoyama et. al. \citep{Isoyama:2013yor,Isoyama:2018sib}, and recently by Nasipak and Evans \citep{Nasipak:2021qfu}, that the evolution of the Carter constant is crucial in order to describe the adiabatic evolution of EMRIs through resonances. Therefore, it could be interesting to include it in the case of electromagnetic case as well, but, we leave this for a future work.

The above discussed discrepancies have as a result that the evolution from the same initial conditions are not giving the same time $\Delta t_r$ that the inspiral spends in the resonance for the adiabatic and the self-force approach (Fig.~\ref{fig:ResTime}). In order to obtain $\Delta t_r$ for different initial conditions, we follow the prescription shown in Ref. \cite{LG21}. Given an initial condition, we evolve the dissipative system for a sufficiently long time to obtain $\sim$ $10^3$ to $10^4$ points in Poincar\'{e} section. By referring to Fig.~\ref{fig:KBH_attractive_dissip}, we encounter similar structure in the Poincar\'{e} section which hints where and when the particle meets the resonance. Once we pin down the locations of the resonance, we note the coordinates of these points, along with the corresponding 4-momentum, energy, angular momentum and the value of the Hamiltonian. Afterwords, we evolve each of these points conservatively for a significant amount of time and evaluate the rotation number. We repeat this procedure for each initial condition given in Fig.~\ref{fig:ResTime}. By looking at the values of $\Delta t_r$ shown in Fig.~\ref{fig:ResTime} it appears that the time spent by the inspiral in the resonance to be qualitatively the same for both approaches. Even exotic cases, like those that the inspiral enters a resonance, but does not leave, seems to be reproduced both by a self-force and an adiabatic evolution. Note that this exotic effect is similar with the sustained resonances \cite{haberman1983energy} appearing in EMRI studies \cite{vandeMeent14}. Hence, the adiabatic scheme appear to be sufficiently faithful to the instantaneous self-force evolution. 

In Fig.~\ref{fig:ResTime} the initial conditions giving the \enquote{sustained} type of resonance crossings are indicated by two nearby vertical lines. These lines lie between a maximum and a minimum of the $\Delta t_r(r)$ plot. The absence of these lines in the self-force evolution scheme through the $1:3$ resonance (left plot of Fig.~\ref{fig:ResTime}) is probably caused by its very small width. The minima of the $\Delta t_r(r)$ plot correspond to inspirals crossing through the vicinity of the unstable periodic orbit of the resonance. The maxima of the $\Delta t_r(r)$ plot correspond to inspirals entering sufficiently deep into the islands of stability (formed around the stable periodic orbit), which spend a considerable time period within the island before they exit the resonance. On the other hand, if an inspiral enters too deep into the island of stability, it becomes trapped by the resonance for a very long time which exceeds the integration time.

\section{Astrophysical relevance of the parameters}\label{sec:Param}

The model was discussed in geometrized units scaled by the rest mass of the central black hole. In order to check the astrophysical consistency of the employed values we employ the relation between radiation parameter $k$, mass ratio $\eta$ and specific charge $\tilde{q}$ given by Eq.~(\ref{radiation_parameter}). In particular, for our numerical examples we employ the value $k=10^{-3}M$, for which the mass ratio yields $\eta \sim 10^{-1}(\tilde{q})^{-2}$. Fixing the mass ratio at the value relevant for EMRI systems as $\eta=10^{-4}$ thus leads to $\tilde{q}\sim 1$. Generally, the upper limit on the relevant value of the specific charge would be set by an electron with $|\tilde{q}_e|\sim10^{21}$ in geometrized units. On the other hand, the theoretical limit on the charge of the static (Reissner-Nordstr\"{o}m) black hole is $|\tilde{q}_{\rm RN}|=1$. For the value employed in our analysis, $|\tilde{\epsilon}|=|\tilde{q}B_0|=10^{-3}M^{-1}$, we may then retrieve the value of magnetic induction in physical units if M is specified:
\begin{equation}
(B)_{SI}=\frac{\tilde{\epsilon}}{\tilde{q}}\frac{c}{1472}\left(\frac{M_{\odot}}{M}\right)\sqrt{\frac{k_C}{G}}.
\label{mag_physical}
\end{equation}
In particular, for a black hole in the center of M87 galaxy with mass $M\sim10^{10}\,M_{\odot}$ \citep{M87_mass} we obtain $(B)_{SI}\sim10^2\,T$. This is several orders of magnitude more than the value derived from the recent observations of M87 with the Event Horizon Telescope \citep{M87_mag}, however,  still within the range of realistic estimates for accreting black holes \citep{sarkar18,delsanto13}.

Dissipative trajectories studied in the present paper were evolved by Eq.~(\ref{eq:ESF_Charged}) in which the contribution of the tail term was neglected. While the estimates presented in \cite{tursunov18} justify such approximation for the case of a single charged particle such as electron near a magnetized black hole, we need to verify its validity for our scenario of an EMRI analogue. To proceed, we employ results from \cite{smith80}, where the self-force on the static point charge $q$ of mass $m$ near a (Schwarzschild) black hole of mass $M$ is computed. In such system with no external electromagnetic field, the self-force appears solely due to the interaction between the field of the point charge and the black hole curvature and thus allows us to estimate the contribution of the tail term (the only part of the self-force which remains when the magnetic field is switched off in our model). The ratio $\Psi$ between the self-force and the gravitational force (which remains dominant in our case as we set $|\epsilon| =10^{-3}$), is shown \citep{smith80} to have its maximum close to horizon (namely at $r=3\,M$ in Schwarzschild spacetime) and drops as $\Psi\propto 1/r$ farther from the black hole. In particular, the maximum value $\Psi^{\rm max}$ (expressed by quantities in SI units) is given as:
\begin{equation}
\Psi^{\rm max}=\frac{k_C}{3\sqrt{3}\,G}\left(\frac{q^2}{m\,M}\right)_{SI}=\frac{k_C\, (\tilde{q})_{SI}^2\,\eta}{3\sqrt{3}\,G},
\label{force_ratio}
\end{equation}
where $k_{C}$ and $G$ are he Coulomb and the gravitational constants, respectively.

For an electron and a black hole of one stellar mass we get $\Psi^{\rm max}\sim10^{-19}$. For the most unfavourable EMRI case of $\eta=10^{-4}$ and $\tilde{q}=1$ (extremal Reissner-Nordstr\"{o}m black hole) the ratio yields $\Psi^{\rm max}\sim10^{-5}$ which makes this contribution negligible even in this worst-case scenario, while for radii corresponding to our numerical examples this ratio reduces at least to $\Psi^{\rm max}\sim10^{-6}$.

The above analysis shows that our model and employed approximations are generally consistent with the conditions encountered in astrophysical systems. However, we stress that it is not proposed as a model directly corresponding to an EMRI and, in particular, the values of mass ratio formally expressed in Eq.~(\ref{radiation_parameter}) cannot be straightforwardly identified with the mass ratio parameter in EMRI system driven by GSF. Instead of modelling particular dynamic properties of an EMRI, the motivation of our analogue model is more general and our aim is to study fundamental properties of resonances affected by a non-integrable perturbation and the behavior of trajectories crossing such resonances due to dissipation caused by a self-force. Our setup allows us to test the reliability of the adiabatic approximation. In particular, in the present work we raised (and positively answered) the question whether the evolution of resonance-crossing trajectories might be reasonably approximated by the adiabatic (averaged) prescription for the dissipation of $\tilde{E}$ and $\tilde{L}_{z}$.

\section{Summary and Discussion} \label{sec:Conc}

In this work we studied the dynamics of a charged body orbiting a magnetized Kerr black hole. This non-integrable system bears some dynamical similarities with the system of a spinning body moving in the pure Kerr background. In particular, the trajectories in both systems deviate from the geodesics: in the first system, this is due to Lorentz force, while in the second due to spin-curvature coupling. In both systems, the induced perturbation breaks the full integrability. In the first case, it is the presence of the magnetic field, while in the second, it is the spin of the secondary body, which makes the system non-integrable. In both systems there is a Carter-like constant, which holds up to linear order in the perturbation term and effects of non-integrability appear due to terms quadratic in the perturbation. This fact has recently been demonstrated for the spinning body \citep{Witzany19,Zelenka20}, while for the charged body orbiting a magnetized Kerr black hole we showed that in \sect{sec:ChargedPartIntro} of the present paper. The above reasons make the latter system an interesting {\em electromagnetic analogue} of an EMRI, which allows to study the dynamics of the inspiralling body during the resonance crossing induced by the self-force.   

In our study we induced dissipation to the charged body using two approaches. First, we considered the instantaneous electromagnetic self-force without its tail terms. We evolved the system through a $1:3$ and $1:2$ resonances and studied the crossings of these resonances for various initial conditions. During the evolution of these crossings, we computed losses of the energy, the angular momentum along $z$ and the Carter-like quantity $\tilde{\mathcal{K}}$. We noticed that although the energy and the angular momentum were changing relatively smoothly, $\tilde{\mathcal{K}}$ experienced an abrupt change due to resonance crossing. It is not clear why only $\tilde{\mathcal{K}}$ (and not the other constants) exhibits such behavior. Is it a feature of the electromagnetic self-force, which will not be reproduced in the GSF case?  Further investigation is needed to determine the reason of this discrepancy.

Since we calculated how the energy and angular momentum change along each trajectory, we were able to interpolate the time evolution of these quantities. Using these interpolations, we applied an adiabatic scheme to evolve orbits crossing $1:3$ and $1:2$ resonances. This allowed us to test whether the adiabatic scheme represents a faithful approximation of an instantaneous self-force. We were able to check that the adiabatic crossings of the resonance last for time intervals that are quantitatively comparable with those given by the instantaneous self-force. This does not mean that there are not discrepancies, like the presence of \enquote{sustained} resonances in the case of the $1:3$ resonance, which occur in the adiabatic approximation, but not with the full self-force. However, this discrepancy might be an artifact of our adiabatic scheme in which the dissipation of the Carter-like constant is not prescribed. We plan to further investigate this issue and possibly optimize our adiabatic approximation.

Nevertheless, the fact that we got faithful results regarding the resonance crossing duration without prescribing the dissipation of the Carter-like constant is remarkable and might also have an application in non-integrable systems where a Carter-like constant does not exist even to a linear order with respect to the perturbation. Our results strongly indicate that it is sufficient to adiabatically dissipate the system just through the energy and angular momentum, in order to find the correct times that resonance crossings last in EMRIs.

Since we studied an analogue model driven by an electromagnetic self-force, any particular numerical values of the observable quantities (e.g., time intervals spent in  resonances) are not directly relevant from the observational perspective of EMRIs. However, the main result of our analysis, which is the remarkable correspondence between the instantaneous self-force and its adiabatic approximation, is supposed to hold for a significantly broader class of non-integrable dynamical systems with dissipation. In particular, our results provide an indication that adiabatic approximation might be sufficient to faithfully model intricate dynamics of resonance crossing of an EMRI.

\begin{acknowledgments}
The authors have been supported by the fellowship Lumina Quaeruntur No. LQ100032102 of the Czech Academy of Sciences. We thank Dr. Martin Kolo\v{s} and Dr. Arman Tursunov for the fruitful discussions.
\end{acknowledgments}
\appendix

\section{Adiabatic approximation}

\begin{figure}[htp]
\vspace{.5cm}
\includegraphics[width=0.95\linewidth]{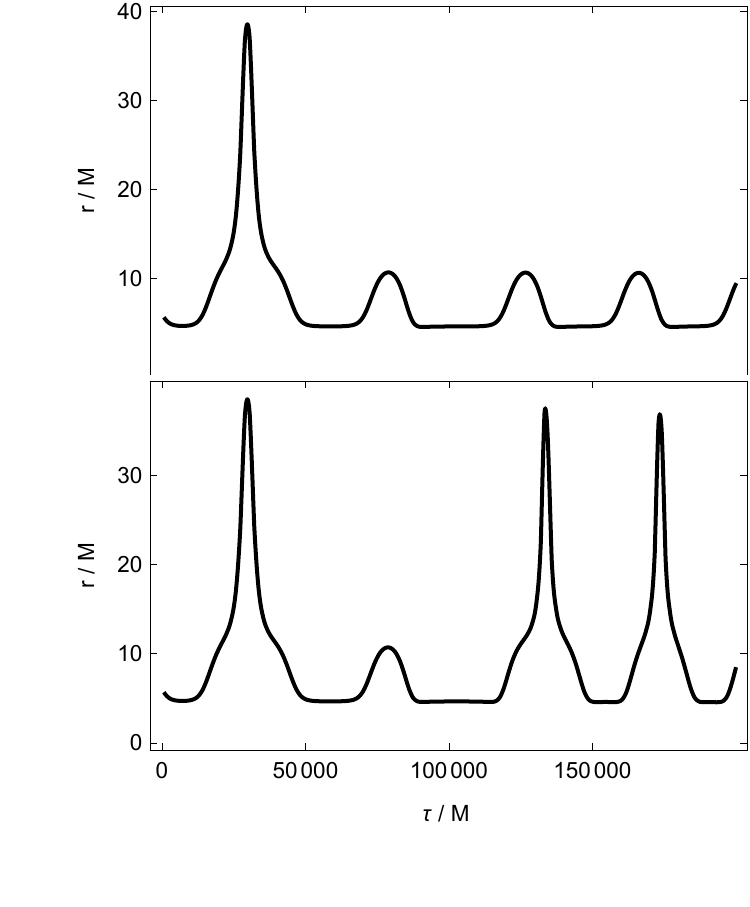}
\caption{ This figure shows how higher order terms in the fitted energy and angular momentum affect the resonance crossing. For the upper panel, we set $ E(\tau)=E_0+a_1 \tau+a_2 \tau^2$, while for the lower pane, we set $E(\tau)=E_0+a_1 \tau+a_2 \tau^2+a_3 \tau^3$. The upper case, the body seems to repeat the loop several times, while for the lower case, the body only cross it once. This example is for the initial distance $r=40.09966M$, $\tilde{E}(0)=0.98$, $\tilde{L_{\rm z}}(0)=3.8M$, $a=0.5M$, $\epsilon=-10^{-3}$, and $k=10^{-3}$.}
\label{fig:poly_fit_energy}
\end{figure}

\begin{figure}[h]
\includegraphics[width=0.95\linewidth]{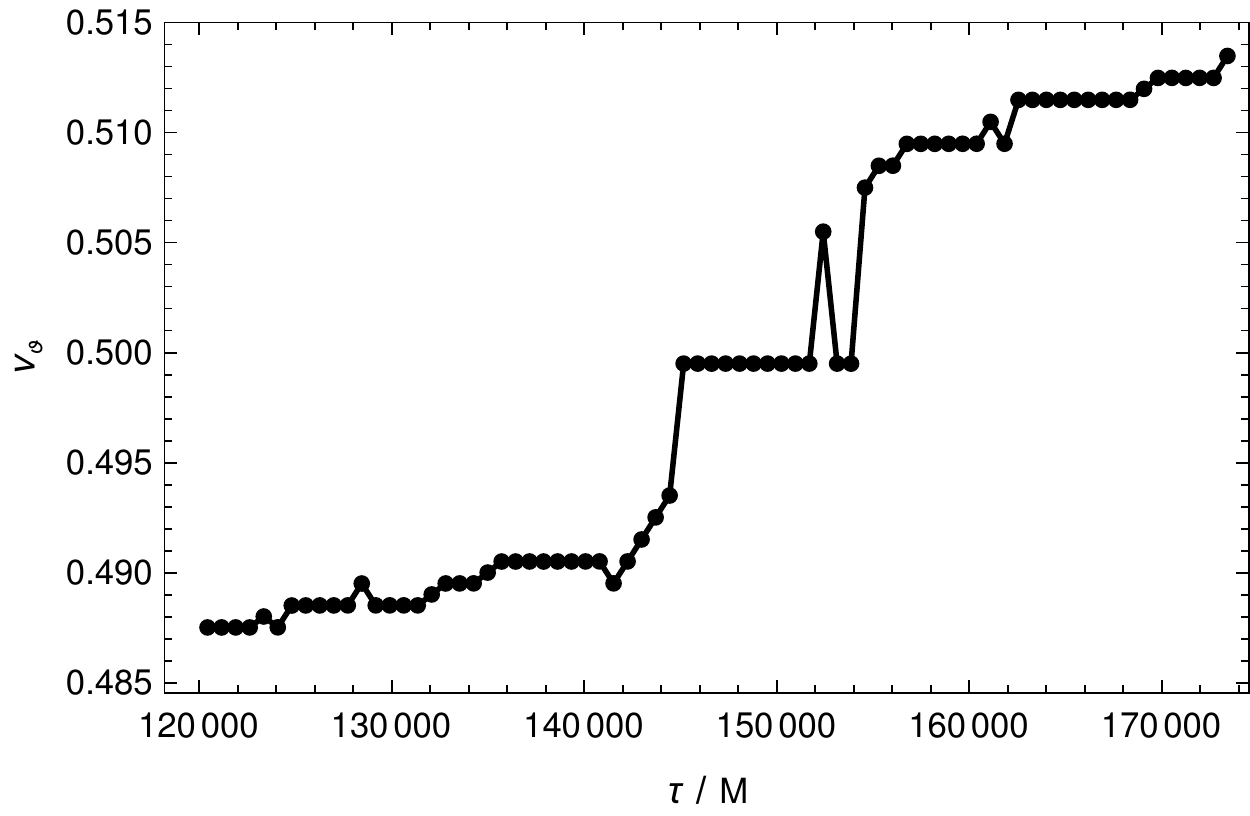}
\caption{The rotation curve for a dissipative system as it inspirals through $1:2$ resonance. The initial conditions are: $\tilde{E}(0)=0.98$, $\tilde{L_{\rm z}}(0)=3.8M$, $a=0.5M$, $\epsilon=-10^{-3}$, $k=10^{3}$, $r=41.19205M$ and $\theta=\pi/2$.}
\label{fig:Deflection}
\end{figure}

Let us now point out a few artifacts of the adiabatic approximation employed in our work. We found that close to the change from the maxima to the minima in Fig.~\ref{fig:ResTime} the adiabatic framework begins to encounter a problematic behavior. 

We demonstrate in Fig.~\ref{fig:poly_fit_energy} how the polynomial fit of energy and angular momentum may affect the adiabatic evolution. Fig.~\ref{fig:poly_fit_energy} shows the radial coordinate of every third crossing of the inspiral through the equatorial plane when $\pi^\theta>0$ as a function of the proper time $\tau$. Basically, we use a stroboscopic depiction of a Poincar\'{e} section as was discussed when introducing the top panel of Fig.~\ref{fig:KBH_attractive_dissip}. The orbit starts on the equatorial plane with $\pi^r=0$ and $r=40.09966M$, but we use for aesthetic reasons the immediately next crossing through the Poincar\'{e} section to produce the stroboscopic picture in both the top panel of Fig.~\ref{fig:KBH_attractive_dissip} and in Fig.~\ref{fig:poly_fit_energy}.  In the latter figure high peaks correspond to the part of the inspiral moving on KAMs away from the resonance, while low peaks correspond to the phase of the evolution spend in the resonance on an island of stability (see top panel Fig.~\ref{fig:KBH_attractive_dissip}). To reproduce the top panel of Fig.~\ref{fig:poly_fit_energy}, we use a second order polynomial fit for the energy and the angular momentum. The multiple low peaks indicate that the inspiral spend a significant number of cycles in the resonance. However, this feature simply disappears if we consider a higher order polynomial fit, as depicted in the lower panel of Fig.~\ref{fig:poly_fit_energy}. Hence, as we move closer and closer to the jump from maxima to minima, we need higher order polynomial fit to avoid such artifacts. Nonetheless, for some initial conditions, as shown within the dashed blue lines of Fig. \ref{fig:ResTime}, it is not possible to avoid inspirals being trapped in the resonance. This kind of entrapment seems to be a feature of the system, since we can see it happening also in the self-force driven evolution. 

For the second example, we again consider the adiabatic approximation close to the jump from maxima to minima. In Fig.~\ref{fig:Deflection}, we provide an example of the adiabatic evolution such that it crosses a $1:2$ resonance. We notice that the journey through the resonance is not smooth, and some points deviate from this plateau. This is not an numerical artifact and not present for the full self-force computations. However, we observe this feature for both $1:2$ and $1:3$ resonances when evolving  the system adiabatically. 

\bibliographystyle{utphys1.bst}
\bibliography{ref}

\providecommand{\href}[2]{#2}\begingroup\raggedright\begin{thebibliography}{10}

\bibitem{lichtenberg92}
A.~{Lichtenberg} and M.~{Lieberman}, {\em {Regular and Chaotic Dynamics}}.
\newblock Springer New York, NY, 1992.

\bibitem{Arnold63}
V.~I. {Arnold}, ``{Proof of a Theorem of A. N. KOLMOGOROV on the Invariance of
  Quasi-Periodic Motions Under Small Perturbations of the Hamiltonian},''
  \href{http://dx.doi.org/10.1070/RM1963v018n05ABEH004130}{{\em Russian
  Mathematical Surveys} {\bfseries 18} no.~5, (Oct., 1963) 9--36}.

\bibitem{Apostolatos09}
T.~A. {Apostolatos}, G.~{Lukes-Gerakopoulos}, and G.~{Contopoulos}, ``{How to
  Observe a Non-Kerr Spacetime Using Gravitational Waves},''
  \href{http://dx.doi.org/10.1103/PhysRevLett.103.111101}{{\em \prl} {\bfseries
  103} no.~11, (Sept., 2009) 111101},
  \href{http://arxiv.org/abs/0906.0093}{{\ttfamily arXiv:0906.0093 [gr-qc]}}.

\bibitem{LGAC10}
G.~Lukes-Gerakopoulos, T.~A. Apostolatos, and G.~Contopoulos, ``{Observable
  signature of a background deviating from the {K}err metric},''
  \href{http://dx.doi.org/10.1103/PhysRevD.81.124005}{{\em Phys. Rev. D}
  {\bfseries 81} (2010) 124005},
  \href{http://arxiv.org/abs/1003.3120}{{\ttfamily arXiv:1003.3120 [gr-qc]}}.

\bibitem{Destounis20}
K.~Destounis, A.~G. Suvorov, and K.~D. Kokkotas, ``Testing spacetime symmetry
  through gravitational waves from extreme-mass-ratio inspirals,''
  \href{http://dx.doi.org/10.1103/PhysRevD.102.064041}{{\em Phys. Rev. D}
  {\bfseries 102} (Sep, 2020) 064041}.
  \url{https://link.aps.org/doi/10.1103/PhysRevD.102.064041}.

\bibitem{Destounis21}
K.~{Destounis}, A.~G. {Suvorov}, and K.~D. {Kokkotas}, ``{Gravitational Wave
  Glitches in Chaotic Extreme-Mass-Ratio Inspirals},''
  \href{http://dx.doi.org/10.1103/PhysRevLett.126.141102}{{\em Phys. Rev.
  Letters} {\bfseries 126} no.~14, (Apr., 2021) 141102},
  \href{http://arxiv.org/abs/2103.05643}{{\ttfamily arXiv:2103.05643 [gr-qc]}}.

\bibitem{LG21}
G.~{Lukes-Gerakopoulos} and V.~{Witzany},
  \href{http://dx.doi.org/10.1007/978-981-15-4702-7_42-1}{``{Nonlinear Effects
  in EMRI Dynamics and Their Imprints on Gravitational Waves},''} in {\em
  Handbook of Gravitational Wave Astronomy}, p.~42.
\newblock 2021.

\bibitem{Bronicki22}
D.~{Bronicki}, A.~{C{\'a}rdenas-Avenda{\~n}o}, and L.~C. {Stein},
  ``{Tidally-induced nonlinear resonances in EMRIs with an analogue model},''
  {\em arXiv e-prints} (Mar., 2022) arXiv:2203.08841,
  \href{http://arxiv.org/abs/2203.08841}{{\ttfamily arXiv:2203.08841 [gr-qc]}}.

\bibitem{EMRIsLISA}
S.~{Babak}, J.~{Gair}, A.~{Sesana}, E.~{Barausse}, C.~F. {Sopuerta}, C.~P.~L.
  {Berry}, E.~{Berti}, P.~{Amaro-Seoane}, A.~{Petiteau}, and A.~{Klein},
  ``{Science with the space-based interferometer LISA. V. Extreme mass-ratio
  inspirals},'' \href{http://dx.doi.org/10.1103/PhysRevD.95.103012}{{\em \prd}
  {\bfseries 95} no.~10, (May, 2017) 103012},
  \href{http://arxiv.org/abs/1703.09722}{{\ttfamily arXiv:1703.09722 [gr-qc]}}.

\bibitem{Barack19}
L.~{Barack} and A.~{Pound}, ``{Self-force and radiation reaction in general
  relativity},'' \href{http://dx.doi.org/10.1088/1361-6633/aae552}{{\em Reports
  on Progress in Physics} {\bfseries 82} no.~1, (Jan., 2019) 016904},
  \href{http://arxiv.org/abs/1805.10385}{{\ttfamily arXiv:1805.10385 [gr-qc]}}.

\bibitem{Pound21}
A.~{Pound} and B.~{Wardell}, ``{Black hole perturbation theory and
  gravitational self-force},'' {\em arXiv e-prints} (Jan., 2021)
  arXiv:2101.04592, \href{http://arxiv.org/abs/2101.04592}{{\ttfamily
  arXiv:2101.04592 [gr-qc]}}.

\bibitem{vandeMeent18}
M.~{van de Meent}, ``{Gravitational self-force on generic bound geodesics in
  Kerr spacetime},'' \href{http://dx.doi.org/10.1103/PhysRevD.97.104033}{{\em
  \prd} {\bfseries 97} no.~10, (May, 2018) 104033},
  \href{http://arxiv.org/abs/1711.09607}{{\ttfamily arXiv:1711.09607 [gr-qc]}}.

\bibitem{Drasco06}
S.~{Drasco} and S.~A. {Hughes}, ``{Gravitational wave snapshots of generic
  extreme mass ratio inspirals},''
  \href{http://dx.doi.org/10.1103/PhysRevD.73.024027}{{\em \prd} {\bfseries 73}
  no.~2, (Jan., 2006) 024027},
  \href{http://arxiv.org/abs/gr-qc/0509101}{{\ttfamily arXiv:gr-qc/0509101
  [gr-qc]}}.

\bibitem{Isoyama21}
S.~{Isoyama}, R.~{Fujita}, A.~J.~K. {Chua}, H.~{Nakano}, A.~{Pound}, and
  N.~{Sago}, ``{Adiabatic waveforms from extreme-mass-ratio inspirals: an
  analytical approach},'' {\em arXiv e-prints} (Nov., 2021) arXiv:2111.05288,
  \href{http://arxiv.org/abs/2111.05288}{{\ttfamily arXiv:2111.05288 [gr-qc]}}.

\bibitem{Katz21}
M.~L. {Katz}, A.~J.~K. {Chua}, L.~{Speri}, N.~{Warburton}, and S.~A. {Hughes},
  ``{Fast extreme-mass-ratio-inspiral waveforms: New tools for millihertz
  gravitational-wave data analysis},''
  \href{http://dx.doi.org/10.1103/PhysRevD.104.064047}{{\em \prd} {\bfseries
  104} no.~6, (Sept., 2021) 064047},
  \href{http://arxiv.org/abs/2104.04582}{{\ttfamily arXiv:2104.04582 [gr-qc]}}.

\bibitem{Flanagan12}
{\'E}.~{\'E}. {Flanagan} and T.~{Hinderer}, ``{Transient Resonances in the
  Inspirals of Point Particles into Black Holes},''
  \href{http://dx.doi.org/10.1103/PhysRevLett.109.071102}{{\em \prl} {\bfseries
  109} no.~7, (Aug., 2012) 071102},
  \href{http://arxiv.org/abs/1009.4923}{{\ttfamily arXiv:1009.4923 [gr-qc]}}.

\bibitem{Brink15}
J.~{Brink}, M.~{Geyer}, and T.~{Hinderer}, ``{Orbital Resonances Around Black
  Holes},'' \href{http://dx.doi.org/10.1103/PhysRevLett.114.081102}{{\em \prl}
  {\bfseries 114} no.~8, (Feb., 2015) 081102},
  \href{http://arxiv.org/abs/1304.0330}{{\ttfamily arXiv:1304.0330 [gr-qc]}}.

\bibitem{Brink:2015roa}
J.~Brink, M.~Geyer, and T.~Hinderer, ``{Astrophysics of resonant orbits in the
  Kerr metric},'' \href{http://dx.doi.org/10.1103/PhysRevD.91.083001}{{\em
  Phys. Rev. D} {\bfseries 91} no.~8, (2015) 083001},
  \href{http://arxiv.org/abs/1501.07728}{{\ttfamily arXiv:1501.07728 [gr-qc]}}.

\bibitem{Mukherjee:2019jhd}
S.~Mukherjee and S.~Tripathy, ``{Resonant orbits for a spinning particle in
  Kerr spacetime},'' \href{http://dx.doi.org/10.1103/PhysRevD.101.124047}{{\em
  Phys. Rev. D} {\bfseries 101} no.~12, (2020) 124047},
  \href{http://arxiv.org/abs/1905.04061}{{\ttfamily arXiv:1905.04061 [gr-qc]}}.

\bibitem{Berry16}
C.~P.~L. {Berry}, R.~H. {Cole}, P.~{Ca{\~n}izares}, and J.~R. {Gair},
  ``{Importance of transient resonances in extreme-mass-ratio inspirals},''
  \href{http://dx.doi.org/10.1103/PhysRevD.94.124042}{{\em \prd} {\bfseries 94}
  no.~12, (Dec., 2016) 124042},
  \href{http://arxiv.org/abs/1608.08951}{{\ttfamily arXiv:1608.08951 [gr-qc]}}.

\bibitem{sarkar21}
A.~{Sarkar}, A.~{Ali}, and S.~{Nasri}, ``{Perturbative correction terms to
  electromagnetic self-force due to metric perturbation: astrophysical and
  cosmological implications},''
  \href{http://dx.doi.org/10.1140/epjc/s10052-021-09485-y}{{\em European
  Physical Journal C} {\bfseries 81} no.~8, (Aug., 2021) 725},
  \href{http://arxiv.org/abs/2008.06095}{{\ttfamily arXiv:2008.06095 [gr-qc]}}.

\bibitem{Zimmerman:2014uja}
P.~Zimmerman and E.~Poisson, ``{Gravitational self-force in nonvacuum
  spacetimes},'' \href{http://dx.doi.org/10.1103/PhysRevD.90.084030}{{\em Phys.
  Rev. D} {\bfseries 90} no.~8, (2014) 084030},
  \href{http://arxiv.org/abs/1406.5111}{{\ttfamily arXiv:1406.5111 [gr-qc]}}.

\bibitem{poisson04}
E.~{Poisson}, ``{The Motion of Point Particles in Curved Spacetime},''
  \href{http://dx.doi.org/10.12942/lrr-2004-6}{{\em Living Reviews in
  Relativity} {\bfseries 7} no.~1, (May, 2004) 6},
  \href{http://arxiv.org/abs/gr-qc/0306052}{{\ttfamily arXiv:gr-qc/0306052
  [gr-qc]}}.

\bibitem{Pound05}
A.~{Pound}, E.~{Poisson}, and B.~G. {Nickel}, ``{Limitations of the adiabatic
  approximation to the gravitational self-force},''
  \href{http://dx.doi.org/10.1103/PhysRevD.72.124001}{{\em \prd} {\bfseries 72}
  no.~12, (Dec., 2005) 124001},
  \href{http://arxiv.org/abs/gr-qc/0509122}{{\ttfamily arXiv:gr-qc/0509122
  [gr-qc]}}.

\bibitem{Pound08}
A.~{Pound} and E.~{Poisson}, ``{Multiscale analysis of the electromagnetic
  self-force in a weak gravitational field},''
  \href{http://dx.doi.org/10.1103/PhysRevD.77.044012}{{\em \prd} {\bfseries 77}
  no.~4, (Feb., 2008) 044012}, \href{http://arxiv.org/abs/0708.3037}{{\ttfamily
  arXiv:0708.3037 [gr-qc]}}.

\bibitem{wald74}
R.~M. {Wald}, ``{Black hole in a uniform magnetic field},''
  \href{http://dx.doi.org/10.1103/PhysRevD.10.1680}{{\em \prd} {\bfseries 10}
  no.~6, (Sept., 1974) 1680--1685}.

\bibitem{bicak85}
J.~{Bicak} and V.~{Janis}, ``{Magnetic fluxes across black holes},''
  \href{http://dx.doi.org/10.1093/mnras/212.4.899}{{\em Monthly Notices of the
  Royal Astronomical Society} {\bfseries 212} (Feb., 1985) 899--915}.

\bibitem{karas13}
V.~{Karas}, O.~{Kop{\'a}{\v{c}}ek}, and D.~{Kunneriath}, ``{Magnetic Neutral
  Points and Electric Lines of Force in Strong Gravity of a Rotating Black
  Hole},'' \href{http://dx.doi.org/10.4236/ijaa.2013.33A003}{{\em International
  Journal of Astronomy and Astrophysics} {\bfseries 3} no.~3, (Jan., 2013)
  18--24}, \href{http://arxiv.org/abs/1303.7251}{{\ttfamily arXiv:1303.7251
  [astro-ph.HE]}}.

\bibitem{balbus98}
S.~A. {Balbus} and J.~F. {Hawley}, ``{Instability, turbulence, and enhanced
  transport in accretion disks},''
  \href{http://dx.doi.org/10.1103/RevModPhys.70.1}{{\em Reviews of Modern
  Physics} {\bfseries 70} no.~1, (Jan., 1998) 1--53}.

\bibitem{kolos21}
M.~{Kolo{\v{s}}}, A.~{Tursunov}, and Z.~{Stuchl{\'\i}k}, ``{Radiative Penrose
  process: Energy gain by a single radiating charged particle in the ergosphere
  of rotating black hole},''
  \href{http://dx.doi.org/10.1103/PhysRevD.103.024021}{{\em \prd} {\bfseries
  103} no.~2, (Jan., 2021) 024021},
  \href{http://arxiv.org/abs/2010.09481}{{\ttfamily arXiv:2010.09481 [gr-qc]}}.

\bibitem{kovar10}
J.~{Kov{\'a}{\v{r}}}, O.~{Kop{\'a}{\v{c}}ek}, V.~{Karas}, and
  Z.~{Stuchl{\'\i}k}, ``{Off-equatorial orbits in strong gravitational fields
  near compact objects{\textemdash}II: halo motion around magnetic compact
  stars and magnetized black holes},''
  \href{http://dx.doi.org/10.1088/0264-9381/27/13/135006}{{\em Classical and
  Quantum Gravity} {\bfseries 27} no.~13, (July, 2010) 135006},
  \href{http://arxiv.org/abs/1005.3270}{{\ttfamily arXiv:1005.3270
  [astro-ph.HE]}}.

\bibitem{kopacek10}
O.~{Kop{\'a}{\v{c}}ek}, V.~{Karas}, J.~{Kov{\'a}{\v{r}}}, and
  Z.~{Stuchl{\'\i}k}, ``{Transition from Regular to Chaotic Circulation in
  Magnetized Coronae near Compact Objects},''
  \href{http://dx.doi.org/10.1088/0004-637X/722/2/1240}{{\em \apj} {\bfseries
  722} no.~2, (Oct., 2010) 1240--1259},
  \href{http://arxiv.org/abs/1008.4650}{{\ttfamily arXiv:1008.4650
  [astro-ph.HE]}}.

\bibitem{alzahrani14}
A.~M. {Al Zahrani}, ``{Escape of charged particles moving around a weakly
  magnetized Kerr black hole},''
  \href{http://dx.doi.org/10.1103/PhysRevD.90.044012}{{\em \prd} {\bfseries 90}
  no.~4, (Aug., 2014) 044012}, \href{http://arxiv.org/abs/1407.7069}{{\ttfamily
  arXiv:1407.7069 [gr-qc]}}.

\bibitem{kopacek18b}
O.~{Kop{\'a}{\v{c}}ek} and V.~{Karas}, ``{Near-horizon Structure of Escape
  Zones of Electrically Charged Particles around Weakly Magnetized Rotating
  Black Hole},'' \href{http://dx.doi.org/10.3847/1538-4357/aaa45f}{{\em \apj}
  {\bfseries 853} no.~1, (Jan., 2018) 53},
  \href{http://arxiv.org/abs/1801.01576}{{\ttfamily arXiv:1801.01576
  [astro-ph.HE]}}.

\bibitem{kopacek20}
O.~{Kop{\'a}{\v{c}}ek} and V.~{Karas}, ``{Near-horizon Structure of Escape
  Zones of Electrically Charged Particles around Weakly Magnetized Rotating
  Black Hole. II. Acceleration and Escape in the Oblique Magnetosphere},''
  \href{http://dx.doi.org/10.3847/1538-4357/ababa8}{{\em \apj} {\bfseries 900}
  no.~2, (Sept., 2020) 119}, \href{http://arxiv.org/abs/2008.04630}{{\ttfamily
  arXiv:2008.04630 [astro-ph.HE]}}.

\bibitem{mtw}
C.~W. {Misner}, K.~S. {Thorne}, and J.~A. {Wheeler}, {\em {Gravitation}}.
\newblock Princeton University Press, 2017.

\bibitem{beskin15}
V.~S. {Beskin}, A.~{Balogh}, M.~{Falanga}, and R.~A. {Treumann}, ``{Magnetic
  Fields at Largest Universal Strengths: Overview},''
  \href{http://dx.doi.org/10.1007/s11214-015-0188-1}{{\em Space Science
  Reviews} {\bfseries 191} no.~1-4, (Oct., 2015) 1--12}.

\bibitem{bicak00}
J.~{Bi{\v{c}}{\'a}k} and T.~{Ledvinka}, ``{Electromagnetic fields around black
  holes and Meissner effect},'' {\em Nuovo Cimento B Serie} {\bfseries 115}
  no.~70809, (July, 2000) 739,
  \href{http://arxiv.org/abs/gr-qc/0012006}{{\ttfamily arXiv:gr-qc/0012006
  [gr-qc]}}.

\bibitem{Gurlebeck17}
N.~{G{\"u}rlebeck} and M.~{Scholtz}, ``{Meissner effect for weakly isolated
  horizons},'' \href{http://dx.doi.org/10.1103/PhysRevD.95.064010}{{\em \prd}
  {\bfseries 95} no.~6, (Mar., 2017) 064010},
  \href{http://arxiv.org/abs/1702.06155}{{\ttfamily arXiv:1702.06155 [gr-qc]}}.

\bibitem{Gurlebeck18}
N.~{G{\"u}rlebeck} and M.~{Scholtz}, ``{Meissner effect for axially symmetric
  charged black holes},''
  \href{http://dx.doi.org/10.1103/PhysRevD.97.084042}{{\em \prd} {\bfseries 97}
  no.~8, (Apr., 2018) 084042},
  \href{http://arxiv.org/abs/1802.05423}{{\ttfamily arXiv:1802.05423 [gr-qc]}}.

\bibitem{kopacek18}
O.~{Kop{\'a}{\v{c}}ek}, T.~{Tahamtan}, and V.~{Karas}, ``{Null points in the
  magnetosphere of a plunging neutron star},''
  \href{http://dx.doi.org/10.1103/PhysRevD.98.084055}{{\em \prd} {\bfseries 98}
  no.~8, (Oct., 2018) 084055},
  \href{http://arxiv.org/abs/1810.04220}{{\ttfamily arXiv:1810.04220
  [astro-ph.HE]}}.

\bibitem{karas12}
V.~{Karas}, O.~{Kop{\'a}{\v{c}}ek}, and D.~{Kunneriath}, ``{Influence of
  frame-dragging on magnetic null points near rotating black holes},''
  \href{http://dx.doi.org/10.1088/0264-9381/29/3/035010}{{\em Classical and
  Quantum Gravity} {\bfseries 29} no.~3, (Feb., 2012) 035010},
  \href{http://arxiv.org/abs/1201.0009}{{\ttfamily arXiv:1201.0009
  [astro-ph.HE]}}.

\bibitem{carter68}
B.~{Carter}, ``{Global Structure of the Kerr Family of Gravitational Fields},''
  \href{http://dx.doi.org/10.1103/PhysRev.174.1559}{{\em Physical Review}
  {\bfseries 174} no.~5, (Oct., 1968) 1559--1571}.

\bibitem{carter1973}
B.~{Carter}, {\em {Black hole equilibrium states, C. deWitt and B. deWitt,
  Eds., Black Holes}}.
\newblock Gordon and Breach Science Publishers, New York, 1973.

\bibitem{Mukherjee:2015oaa}
S.~Mukherjee and K.~Rajesh~Nayak, ``{Carter constant and angular momentum},''
  \href{http://dx.doi.org/10.1142/S0218271817501802}{{\em Int. J. Mod. Phys. D}
  {\bfseries 27} no.~01, (2017) 1750180},
  \href{http://arxiv.org/abs/1507.01863}{{\ttfamily arXiv:1507.01863 [gr-qc]}}.

\bibitem{Morbidelli02}
A.~Morbidelli, {\em Modern celestial mechanics: aspects of solar system
  dynamics}.
\newblock CRC Press, 1st~ed., 2002.

\bibitem{Birkhoff13}
G.~D. Birkhoff, ``Proof of poincaré's geometric theorem,'' {\em Transactions
  of the American Mathematical Society} {\bfseries 14} no.~1, (1913) 14--22.
  \url{http://www.jstor.org/stable/1988766}.

\bibitem{Voglis98}
N.~Voglis and C.~Efthymiopoulos, ``{Angular dynamical spectra. A new method for
  determining frequencies, weak chaos and cantori},''
  \href{http://dx.doi.org/10.1088/0305-4470/31/12/015}{{\em J. Phys. A: Math.
  Gen.} {\bfseries 31} (1998) 2913--2928}.

\bibitem{Contopoulos02}
G.~Contopoulos, {\em Order and chaos in dynamical astronomy}.
\newblock Springer Science \& Business Media, 2002.

\bibitem{Zelenka20}
O.~{Zelenka}, G.~{Lukes-Gerakopoulos}, V.~{Witzany}, and
  O.~{Kop{\'a}{\v{c}}ek}, ``{Growth of resonances and chaos for a spinning test
  particle in the Schwarzschild background},''
  \href{http://dx.doi.org/10.1103/PhysRevD.101.024037}{{\em \prd} {\bfseries
  101} no.~2, (Jan., 2020) 024037},
  \href{http://arxiv.org/abs/1911.00414}{{\ttfamily arXiv:1911.00414 [gr-qc]}}.

\bibitem{rudiger1981conserved}
R.~R{\"u}diger, ``Conserved quantities of spinning test particles in general
  relativity. i,'' {\em Proceedings of the Royal Society of London. A.
  Mathematical and Physical Sciences} {\bfseries 375} no.~1761, (1981)
  185--193.

\bibitem{rudiger1983conserved}
R.~R{\"u}diger, ``Conserved quantities of spinning test particles in general
  relativity. ii,'' {\em Proceedings of the Royal Society of London. A.
  Mathematical and Physical Sciences} {\bfseries 385} no.~1788, (1983)
  229--239.

\bibitem{gibbons1993susy}
G.~Gibbons, R.~H. Rietdijk, and J.~Van~Holten, ``Susy in the sky,'' {\em
  Nuclear Physics B} {\bfseries 404} no.~1-2, (1993) 42--64.

\bibitem{Tanaka:1996ht}
T.~Tanaka, Y.~Mino, M.~Sasaki, and M.~Shibata, ``{Gravitational waves from a
  spinning particle in circular orbits around a rotating black hole},''
  \href{http://dx.doi.org/10.1103/PhysRevD.54.3762}{{\em Phys. Rev.} {\bfseries
  D54} (1996) 3762--3777},
\href{http://arxiv.org/abs/gr-qc/9602038}{{\ttfamily arXiv:gr-qc/9602038
  [gr-qc]}}.

\bibitem{Witzany19}
V.~{Witzany}, ``{Hamilton-Jacobi equation for spinning particles near black
  holes},'' \href{http://dx.doi.org/10.1103/PhysRevD.100.104030}{{\em \prd}
  {\bfseries 100} no.~10, (Nov., 2019) 104030},
  \href{http://arxiv.org/abs/1903.03651}{{\ttfamily arXiv:1903.03651 [gr-qc]}}.

\bibitem{rohrlich2020classical}
F.~Rohrlich, {\em Classical charged particles: foundations of their theory}.
\newblock CRC Press, 2020.

\bibitem{mcdonald2018history}
K.~T. McDonald, ``On the history of the radiation reaction,'' 2018.

\bibitem{Dirac:1938nz}
P.~A.~M. Dirac, ``{Classical theory of radiating electrons},''
  \href{http://dx.doi.org/10.1098/rspa.1938.0124}{{\em Proc. Roy. Soc. Lond. A}
  {\bfseries 167} (1938) 148--169}.

\bibitem{landau1975classical}
L.~D. Landau and E.~Lifshitz, {\em The Classical Theory of Fields: Volume 2},
  vol.~2.
\newblock Butterworth-Heinemann, 1975.

\bibitem{dewitt1960radiation}
B.~S. DeWitt and R.~W. Brehme, ``Radiation damping in a gravitational field,''
  {\em Annals of Physics} {\bfseries 9} no.~2, (1960) 220--259.

\bibitem{tursunov18}
A.~{Tursunov}, M.~{Kolo{\v{s}}}, Z.~{Stuchl{\'\i}k}, and D.~V. {Gal'tsov},
  ``{Radiation Reaction of Charged Particles Orbiting a Magnetized
  Schwarzschild Black Hole},''
  \href{http://dx.doi.org/10.3847/1538-4357/aac7c5}{{\em \apj} {\bfseries 861}
  no.~1, (July, 2018) 2}, \href{http://arxiv.org/abs/1803.09682}{{\ttfamily
  arXiv:1803.09682 [gr-qc]}}.

\bibitem{lorentz1892theorie}
H.~A. Lorentz, {\em La th{\'e}orie {\'e}lectromagn{\'e}tique de Maxwell et son
  application aux corps mouvants}, vol.~25.
\newblock EJ Brill, 1892.

\bibitem{Destounis21b}
K.~{Destounis} and K.~D. {Kokkotas}, ``{Gravitational-wave glitches: Resonant
  islands and frequency jumps in nonintegrable extreme-mass-ratio inspirals},''
  \href{http://dx.doi.org/10.1103/PhysRevD.104.064023}{{\em Phys. Rev. D}
  {\bfseries 104} no.~6, (Sept., 2021) 064023},
  \href{http://arxiv.org/abs/2108.02782}{{\ttfamily arXiv:2108.02782 [gr-qc]}}.

\bibitem{Isoyama:2013yor}
S.~Isoyama, R.~Fujita, H.~Nakano, N.~Sago, and T.~Tanaka, ``{Evolution of the
  Carter constant for resonant inspirals into a Kerr black hole: I. The scalar
  case},'' \href{http://dx.doi.org/10.1093/ptep/ptt034}{{\em PTEP} {\bfseries
  2013} no.~6, (2013) 063E01}, \href{http://arxiv.org/abs/1302.4035}{{\ttfamily
  arXiv:1302.4035 [gr-qc]}}.

\bibitem{Isoyama:2018sib}
S.~Isoyama, R.~Fujita, H.~Nakano, N.~Sago, and T.~Tanaka,
  ``{\textquotedblleft{}Flux-balance formulae\textquotedblright{} for extreme
  mass-ratio inspirals},'' \href{http://dx.doi.org/10.1093/ptep/pty136}{{\em
  PTEP} {\bfseries 2019} no.~1, (2019) 013E01},
  \href{http://arxiv.org/abs/1809.11118}{{\ttfamily arXiv:1809.11118 [gr-qc]}}.

\bibitem{Nasipak:2021qfu}
Z.~Nasipak and C.~R. Evans, ``{Resonant self-force effects in
  extreme-mass-ratio binaries: A scalar model},''
  \href{http://dx.doi.org/10.1103/PhysRevD.104.084011}{{\em Phys. Rev. D}
  {\bfseries 104} no.~8, (2021) 084011},
  \href{http://arxiv.org/abs/2105.15188}{{\ttfamily arXiv:2105.15188 [gr-qc]}}.

\bibitem{haberman1983energy}
R.~Haberman, ``Energy bounds for the slow capture by a center in sustained
  resonance,'' {\em SIAM Journal on Applied Mathematics} {\bfseries 43} no.~2,
  (1983) 244--256.

\bibitem{vandeMeent14}
M.~{van de Meent}, ``{Conditions for sustained orbital resonances in extreme
  mass ratio inspirals},''
  \href{http://dx.doi.org/10.1103/PhysRevD.89.084033}{{\em \prd} {\bfseries 89}
  no.~8, (Apr., 2014) 084033}, \href{http://arxiv.org/abs/1311.4457}{{\ttfamily
  arXiv:1311.4457 [gr-qc]}}.

\bibitem{M87_mass}
E.~H.~T. Collaboration, ``{First M87 Event Horizon Telescope Results. VI. The
  Shadow and Mass of the Central Black Hole},''
  \href{http://dx.doi.org/10.3847/2041-8213/ab1141}{{\em Astrophysical Journal
  Letters} {\bfseries 875} no.~1, (Apr., 2019) L6},
  \href{http://arxiv.org/abs/1906.11243}{{\ttfamily arXiv:1906.11243
  [astro-ph.GA]}}.

\bibitem{M87_mag}
E.~H.~T. Collaboration, ``{First M87 Event Horizon Telescope Results. VIII.
  Magnetic Field Structure near The Event Horizon},''
  \href{http://dx.doi.org/10.3847/2041-8213/abe4de}{{\em Astrophysical Journal
  Letters} {\bfseries 910} no.~1, (Mar., 2021) L13},
  \href{http://arxiv.org/abs/2105.01173}{{\ttfamily arXiv:2105.01173
  [astro-ph.HE]}}.

\bibitem{sarkar18}
B.~{Sarkar} and S.~{Das}, ``{Standing shocks in magnetized dissipative
  accretion flow around black holes},''
  \href{http://dx.doi.org/10.1007/s12036-017-9503-4}{{\em Journal of
  Astrophysics and Astronomy} {\bfseries 39} no.~1, (Feb., 2018) 3},
  \href{http://arxiv.org/abs/1712.07588}{{\ttfamily arXiv:1712.07588
  [astro-ph.HE]}}.

\bibitem{delsanto13}
M.~{Del Santo}, J.~{Malzac}, R.~{Belmont}, L.~{Bouchet}, and G.~{De Cesare},
  ``{The magnetic field in the X-ray corona of Cygnus X-1},''
  \href{http://dx.doi.org/10.1093/mnras/sts574}{{\em Monthly Notices of the
  Royal Astronomical Society} {\bfseries 430} no.~1, (Mar., 2013) 209--220},
  \href{http://arxiv.org/abs/1212.2040}{{\ttfamily arXiv:1212.2040
  [astro-ph.HE]}}.

\bibitem{smith80}
A.~G. {Smith} and C.~M. {Will}, ``{Force on a static charge outside a
  Schwarzschild black hole},''
  \href{http://dx.doi.org/10.1103/PhysRevD.22.1276}{{\em \prd} {\bfseries 22}
  no.~6, (Sept., 1980) 1276--1284}.

\end{thebibliography}\endgroup

\end{document}